\documentclass{aa}  

\usepackage{graphicx}
\usepackage{txfonts}
\usepackage{hyperref}
\begin{document}

   \title{The SOMA-POL Survey. I. Polarization and magnetic field properties of massive protostars}

   \author{Tuva Källberg\inst{1}\fnmsep\thanks{tuva.kallberg@gmail.com},
          Chi Yan Law\inst{2}\fnmsep\inst{3}\fnmsep\thanks{chi.law@inaf.it},
          Jonathan C. Tan\inst{3}\fnmsep\inst{4},  
          Kate Pattle\inst{5}
          \and
          Zacariyya Khan\inst{5}
          }
          
   \institute{KTH Royal Institute of Technology, SE-100 44, Stockholm, Sweden
         \and
            INAF Osservatorio Astrofisico di Arcetri, Largo Enrico Fermi, 5, 50125 Firenze, Italy
         \and 
             Department of Space, Earth \& Environment, Chalmers University of Technology, SE-412 96 Gothenburg, Sweden
        \and 
            Department of Astronomy, University of Virginia, Charlottesville, VA 22904-4235, USA
        \and   
            Department of Physics \& Astronomy, University College London, Gower Street, London, WC1E 6BT, UK.
        }

   \date{}

 
  \abstract{
  The role of magnetic fields in regulating the formation of massive stars remains much debated. Here we present sub-millimeter polarimetric observations with JCMT-POL2 at $850\:\mu$m of 13 regions of massive star formation selected from the SOFIA Massive (SOMA) star formation survey, yielding a total of 29 massive protostars. Our investigation of the $p'-I$ relationship suggests that grain alignment persists up to the highest intensities. We examine the relative orientations between polarization-inferred magnetic field direction and source column density elongation direction on small and large scales. On small scales, we find a bimodal distribution of these relative orientations, i.e., with an excess of near-parallel and near-perpendicular orientations. By applying a one-sample Kuiper test and Monte Carlo simulations to compare to a relative orientation distribution drawn from a uniform distribution, we statistically confirm this bimodal distribution, independent of the methods to measure structural orientation. This bimodal distribution suggests that magnetic fields are dynamically important on the local scales ($\lesssim 0.6\:$pc) of massive protostellar cores. We also examine how basic polarization properties of overall degree of polarization and local dispersion in polarization vector orientations depend on intrinsic protostellar properties inferred from spectral energy distribution (SED) modeling. We find a statistically significant anti-correlation between the debiased polarized fraction and the luminosity to mass ratio, $L_{\rm bol}/M_{\rm env}$, which hints at a change in the dust properties for protostellar objects at different evolutionary stages.}

   
    \titlerunning{SOMA-POL I}
    \authorrunning{T. Källberg et al.}
   \maketitle
%

\section{Introduction}\label{sec:intro}

The importance of magnetic fields in massive star and star cluster formation remains debated. Some theoretical and numerical models assume magnetic fields play an important or dominant role in regulating collapse and fragmentation \citep[e.g.,][]{mckee2003formation,2009MNRAS.399L..94K,2017ApJ...841...88W}. On the other hand, many other such studies, especially those involving competitive accretion, either ignore magnetic fields or assume they are weak and dynamically unimportant \citep[e.g.,][]{2001MNRAS.323..785B,2022MNRAS.512..216G}. Observationally, a number of results have suggested that magnetic ($B-$) fields may be dynamically important in regulating massive star and star cluster formation from cloud ($> 1$ pc) to disk scales ($\sim 100$ au) \citep[e.g.,][]{2009Sci...324.1408G,2014ApJ...792..116Z,li2014link,2015ApJ...799...74P,pattle2022magnetic,law2024polarized}. 
However, some other studies have found more limited evidence for strong magnetic fields \citep[e.g.,][]{2022Natur.601...49C}. 

To further understand the importance of magnetic fields in massive star formation our approach is to conduct a statistical analysis of the $B$-field morphology and related polarization properties in a large sample of massive protostars. In particular, we focus on one of the basic observables, the relative orientations between the average $B$-field direction inferred from dust polarization and the column density structure (hereafter cloud-field alignment). 

While this metric has not yet been investigated in the high-mass regime, using near-infrared dust extinction maps and optical stellar polarimetry data toward 13 Gould Belt molecular clouds, \citet{li2013link} discovered that the molecular cloud major axes tend to be aligned either perpendicular or parallel to the mean magnetic field directions. This bimodal cloud-field alignment was further confirmed with sub-millimeter {\it Planck} column density maps and polarimetry data by \citet{gu2019comparison}. \citet{li2013link} suggested that different cloud-field alignments could originate from different filamentary cloud formation mechanisms. Accumulation of sub-Alfv\'enic turbulent gas along the dynamically important $B$-field lines could lead to a parallel cloud-field alignment, while $B$-field guided compressive contraction could result in perpendicular cloud-field alignment \citep{li2013link,2025arXiv250113931A}. Different cloud-field alignments would cause different magnetic flux, which then affects the corresponding cloud fragmentation properties \citep[e.g.,][]{seifried2015impact,li2017link}. In fact, follow-up simulations and observational studies have demonstrated that the Gould Belt molecular clouds with different cloud-field alignments exhibit different spatial mass distribution ratios, slopes of cumulative mass function distribution, and star formation rates \citep{li2017link,law2019link,law2020links,soler2020history}. \citet{2014ApJ...792..116Z} performed a statistical study on the relative orientation between the core major axis and the local magnetic field orientation and obtained a similar bi-modal distribution, which suggests that magnetic fields continue to play a regulative role down to $0.1~$pc scales.

To study the importance of magnetic fields in massive star formation we have initiated the SOMA-POL survey (PI: C-Y. Law) to obtain $B$-field properties for a large sample of high-mass star-forming regions selected from the SOFIA Massive (SOMA) Star Formation survey (PI: J. C. Tan). The SOMA survey has studied over 50 high- and intermediate-mass star-forming regions across a range of evolutionary stages, masses, and environments with the FORCAST instrument, covering wavelengths ranging from about 7 to 40 $\mu$m. This provides a prime database to statistically study the connection between magnetic field and high-mass star formation. The SOMA-POL survey is a follow-up of the SOMA sample, aimed at studying the polarized dust emission of all sources to constrain the role of $B-$fields in massive star formation and help test different formation theories. In the context of the core accretion model, $B-$fields may provide significant support that helps control fragmentation of protocluster clump gas into a population of cores, i.e., the pre-stellar core mass function, and then may also regulate the rate and geometry of collapse, transfer of angular momentum and disk formation \citep[e.g.,][]{li2014link,tan2014massive}.

In this first paper of the SOMA-POL survey, we report the relative orientation between the mean $B-$field orientation and the orientation of density structures identified in 29 protostars found in 13 SOMA high-mass star-forming regions. As part of this analysis, we also examine how the basic polarization properties of overall degree of polarization and local dispersion in polarization vector orientations depend on intrinsic protostellar properties.

The paper is structured as follows. In \S\ref{sec:obs} we describe the observations and the reduction of the data, including the observed sources and the maps of their Stokes $I$ parameter. In \S\ref{sec:methods} we describe the ways in which the magnetic field orientation and polarization fraction were calculated from the observed data. In \S\ref{sec:SED} we explain how the SED-analysis of the sources was conducted, and also the derivation of the optimal aperture for each source, which is then used for further analysis. In \S\ref{sec:results_analysis} we present basic results, such as magnetic field orientation, angular dispersion and polarization fraction. In \S\ref{sec:analysis} we describe and discuss more advanced analysis, including derivation of structural orientations and statistical methods to study correlations between obtained properties. Finally, the conclusions are presented in \S\ref{sec:conclusion}.
\section{Observations \& data reduction}\label{sec:obs}

\begin{figure*}
    \centering
    \includegraphics[width=1\linewidth]{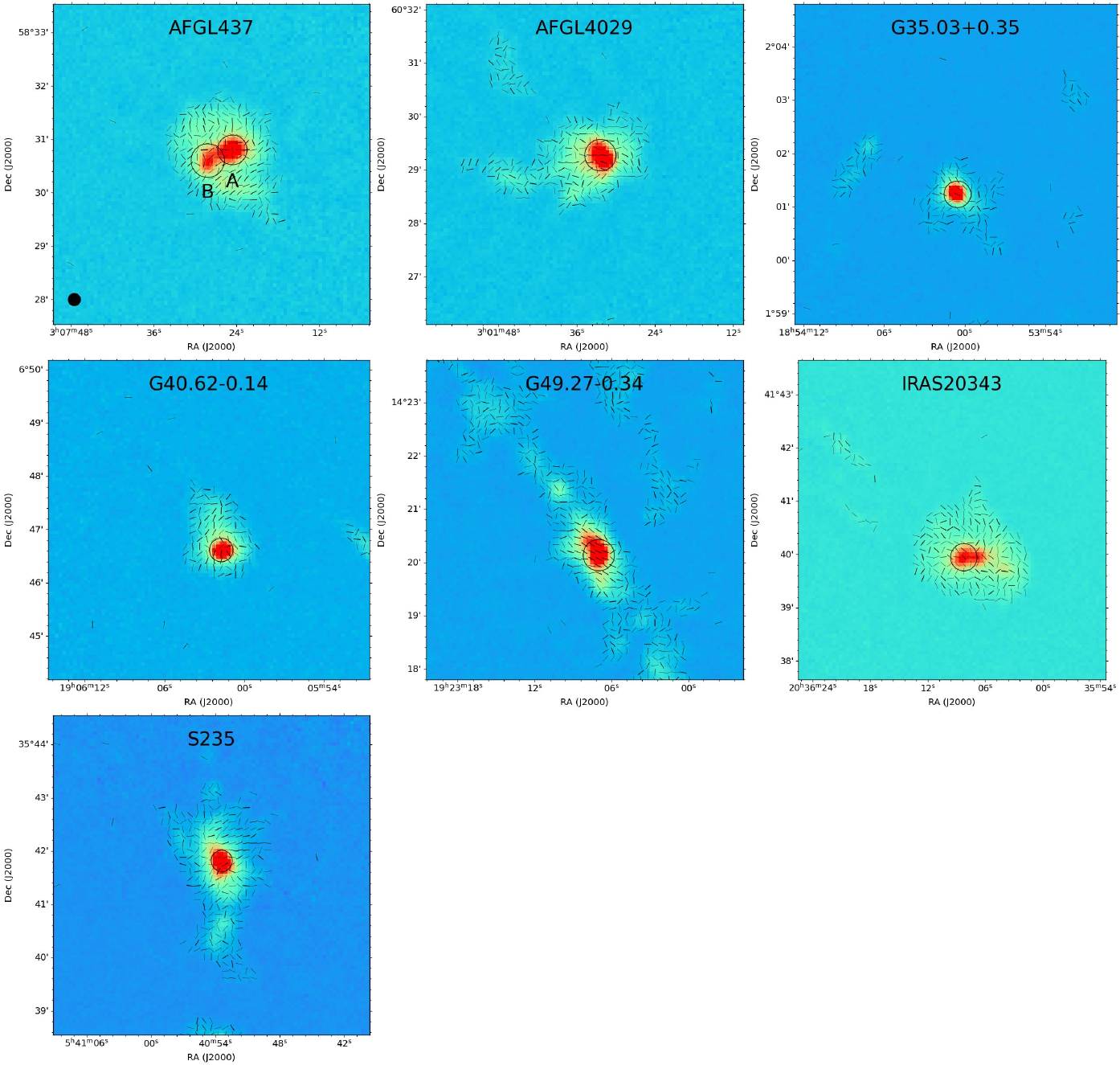}
    \caption{JCMT-POL2 $850\,\mu$ m continuum images of the 7 SOMA regions with one source or binary sources. The sources selected for analysis are marked with black circles with a radius corresponding to the optimal aperture defined in \S\ref{sec:SED}. The black solid circle in the top left panel represents the JCMT beam aperture. Here, all images has a field of view of 3$^\prime$.}
    \label{fig:sources_single}
\end{figure*}

\begin{figure*}
    \centering
    \includegraphics[width=1\linewidth]{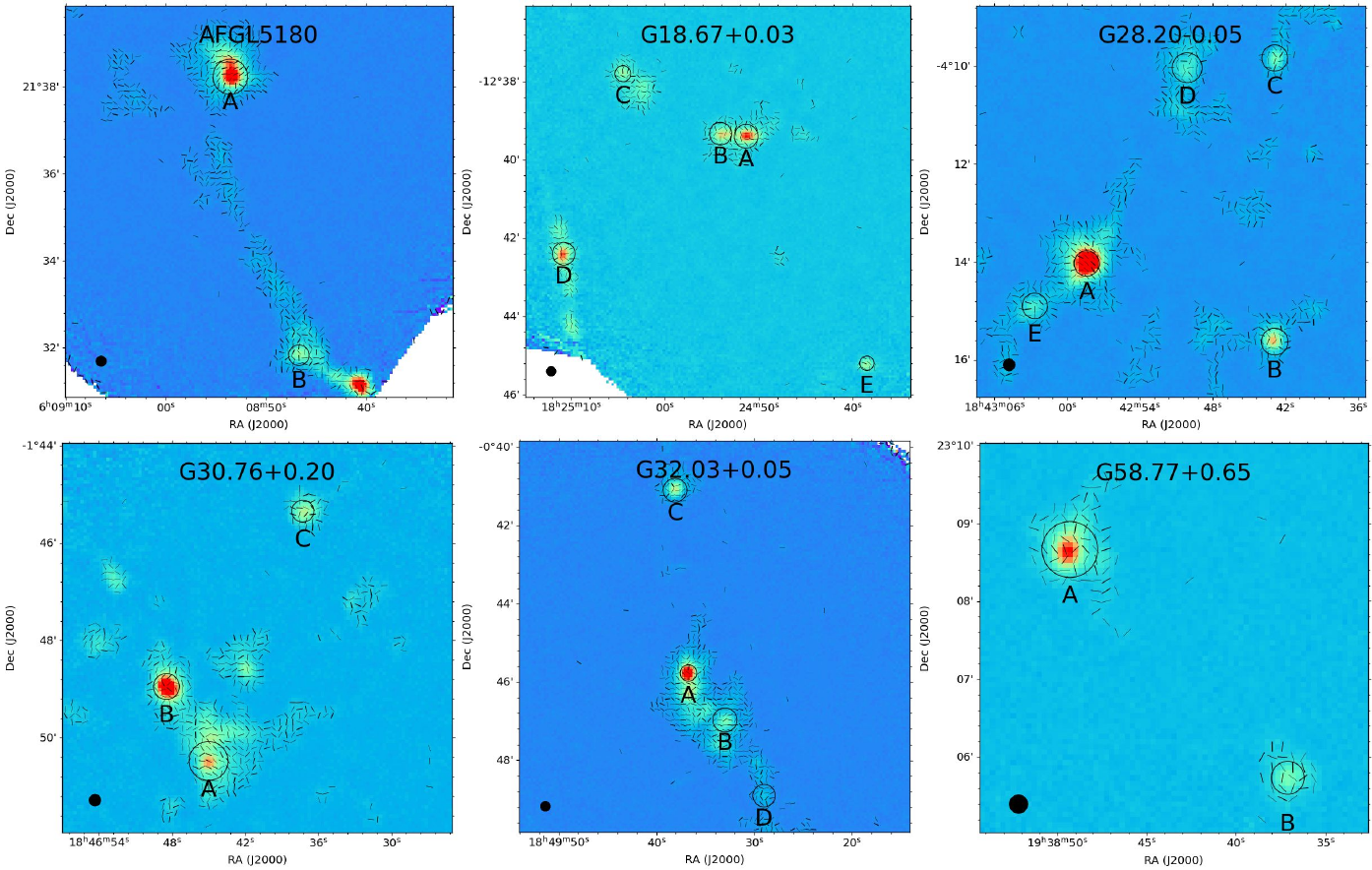}
    \caption{Cont. of Figure. 1. Continuum images of the 6 SOMA regions with more than 2 sources. Black solid circles representing the JCMT beam aperture are here shown in all figures because of the different fields of view.}
    \label{fig:fig1b}
\end{figure*}

We carried out observations of polarized continuum emission at $850\,\mu$m with the JCMT POL-2 camera covering a field-of-view of $12^{\prime}$ by $12^{\prime}$ toward 13 selected SOMA high-mass star-forming regions (see Table \ref{tab:sourceinfo}). These observations were carried out between July 2023 and Jan. 2024. The technical details and data reduction methods are similar to those of the BISTRO survey \citep{doi2020jcmt}. In summary, the reduction was carried out with the {\it starlink} software. The products provide Stokes $I$, $Q$ and $U$ parameters in grid sizes of 4$^{\prime\prime}$, 8$^{\prime\prime}$ and 12$^{\prime\prime}$. In this work, we will use the 4$^{\prime\prime}$ grid for the analysis. We also demonstrate that the mean $B-$field orientations do not change significantly when using different grid sizes (see Appendix A). The $850\,\mu$m continuum images of all regions and sources are displayed in Figure \ref{fig:sources_single}-\ref{fig:fig1b}, with targeted sources for analysis marked with black circles, with a radius corresponding to the optimal (SOMA) aperture for that source (see \S\ref{sec:SED}). The JCMT POL-2 beam aperture of $14^{\prime\prime}$ is also displayed as a solid black circle

\begin{table*}
\centering
\caption{\centering Observed SOMA regions. The table includes coordinates for the center of the region, distance to the region ($d$), observation date, and whether or not we have obtained \textit{Herschel $70\mu m$} data for the region. The center coordinates are based on the $70\mu m$ Herschel data.}
\begin{tabular}{lcccc}
    \hline\hline
    Region & \multicolumn{2}{c}{Coordinates} & $d$ & $70\mu m?$ \\
    \cline{2-3}
     &  R.A. (J2000) &  Decl. (J2000) & (kpc) &   \\
    \hline
    AFGL 437 & $03^h07^m23.9724^s$ & $+58^d30^m52.564^s$ & 2.0 & Y   \\
    \hline
   AFGL 4029 &  $03^h01^m31.2142^s$ & $+60^d29^m12.078^s$ & 2.2 & Y  \\ 
    \hline
    AFGL 5180 & $06^h08^m53.8128^s$ & $+21^d38^m32.533^s$ & 2.2 & Y   \\
    \hline
    G18.67 & $18^h24^m51.0429^s$ & $-12^d39^m21.741^s$ & 11 & Y \\ 
    \hline
    G28.20 & $18^h42^m58.111^s$ & $-4^d13^m57.777^s$ & 5.3 & Y  \\ 
    \hline
    G30.76 & $18^h46^m44.7523^s$ & $-1^d50^m30.533^s$ & 4.9 & Y  \\ 
    \hline
    G32.03 & $18^h49^m36.6808^s$ & $-0^d45^m45.913^s$ & 4.9 & Y  \\ 
    \hline
    G35.03 & $18^h54^m00.5846^s$ & $+2^d01^m18.135^s$ & 3.2 & Y  \\ 
    \hline
   G40.62 & $19^h06^m01.6686^s$ &  $+6^d46^m38.219^s$ & 10.5 & Y  \\ 
    \hline
   G49.27 & $19^h23^m06.6453^s$ & $+14^d20^m10.977^s$ & 5.5 & Y   \\
    \hline
    G58.77 &  $19^h38^m49.1439^s$ & $+23^d08^m39.662^s$ & 3.3 & Y   \\
    \hline
    IRAS 20343 & $20^h36^m07.1385^s$ & $+41^d39^m53.873^s$ & 1.4  & Y  \\
    \hline
    S235$^*$ & $05^h40^m53.4000^s$ & $+35^d41^m49.034^s$ & 1.8 & N   \\
    \hline
\end{tabular}
\tablefoot{Regions marked with $^*$ have coordinates based on the $850 \mu m$ JCMT data instead of the $70\mu m$ Herschel data.}
\label{tab:sourceinfo}
\end{table*}


\section{Polarization angle orientation and polarization fraction}\label{sec:methods}

\subsection{Polarization angle orientation}

The polarization angle $\theta_p$ (in degrees) was obtained from the Stokes $Q$ and Stokes $U$ maps using 
\begin{equation}
    \theta_p = \frac{90}{\pi} \arctan\left( \frac{U}{Q}\right),
    \label{eq:pol_angle}
\end{equation}
with the associated error 
\begin{equation}
    \sigma_{\theta} = \frac{90}{\pi\left(Q^2+U^2\right)}\sqrt{Q^2\sigma_Q +U^2\sigma_U},
    \label{eq:B_angle_error}
\end{equation}
where $\sigma_Q$ and $\sigma_U$ are the errors in Stokes $Q$ and $U$, respectively \citep{datatutorial}. Elongated dust particles tend to align with their major axis perpendicular to the magnetic field, yielding a polarization angle perpendicular to the magnetic field orientation, $\theta_{B}$, \citep{andersson2015interstellar}, i.e., 
\begin{equation}
    \theta_B = \theta_p + 90^\circ.
    \label{eq:B_angle}
\end{equation}

\subsection{Polarization fraction}

The polarization percentage ($p$) was calculated from the Stokes parameters following the equation 
\citep{andersson2015interstellar}:
\begin{equation}
    p = 100 \times \sqrt{\left(\frac{Q}{I}\right)^2 + \left(\frac{U}{I}\right)^2 }
    \label{eq:pol_frac}
\end{equation}
and with the associated error defined by 
\begin{equation}
\begin{split}
\sigma_p = \frac{100}{I} & \left(
\frac{1}{\sqrt{Q^2 + U^2}} \left[ 
(Q \, \sigma_Q)^2 + (U \, \sigma_U)^2 + 2QU \, \sigma_{QU} 
\right] \right.\\
& \left. + \left[ \left( \frac{Q}{I} \right)^2 + \left( \frac{U}{I} \right)^2 \right] \sigma_I^2 
- 2 \frac{Q}{I} \sigma_{QI} - 2 \frac{U}{I} \sigma_{UI}
\right)^{1/2}.
\end{split}
\label{eq:pol_frac error}
\end{equation}

The debiased polarization fraction, $p'$, was then computed by subtracting the error from the polarization fraction via
\begin{equation}
    p' = \sqrt{p^2 - \sigma_p^2}.
    \label{eq:debiased pol_frac}
\end{equation}


\section{Optimal aperture and protostellar properties from SED modeling}\label{sec:SED}

The spectral energy distribution (SED) from near-infrared (NIR) to far-infrared (FIR) can be used to characterize the properties of massive protostars. Here we use the protostellar radiative transfer model grid from \cite{zhang2018radiation} based on the turbulent core accretion model \citep{mckee2002massive,mckee2003formation} to estimate protostellar properties, including current envelope mass $M_{\rm env}$, bolometric luminosity $L_{\rm bol}$, mass surface density of the clump environment $\Sigma_{\rm cl}$ and current mass of the protostar star $m_\star$. To perform the SED analysis, we made use of the python package {\it sedcreator} \citep{fedriani2023}.

We define an optimal (SOMA) aperture for each source based on the \textit{Herschel} $70\mu m$ data using \textit{sedcreator's} get optimal aperture feature. The SOMA aperture is then used for all further analysis in this work. {\it sedcreator} employs an algorithm to systematically identify the aperture. The user defines a lower and upper aperture boundary. Starting from the lower boundary, the aperture radius is increased by 30\% until the resulting increase in background-subtracted flux is $<$10\%, indicating that further increases would yield only a small gain in flux while introducing additional noise. In this work, the lower boundary is set to 3 arcseconds and in general, the upper limit is set to 40 arcseconds. In cases where the sources are more clustered, this upper limit is lowered to reduce the likelihood of including too much flux from secondary sources.

To perform the SED fitting, we use archival {\it Spitzer} and {\it Herschel} data and follow the method described in \citet{fedriani2023} using {\it sedcreator}. Table \ref{tab:SED_params} shows the relevant star formation parameters obtained from the average model of the SED fitting for the source. Specifically, we are interested in the bolometric luminosity and the envelope mass $L_{\rm bol}/M_{\rm env}$ ratio as this indicates the evolutionary stage of the protostar. Results are not provided for G28.2 E as this source is not distinguishable in the Herschel 70$\mu$m data. Values for $S235$ are sourced from \cite{liu2020sofia}.

\begin{table*}
  \centering
  \caption{Estimation of star formation properties from SED analysis: envelope mass, bolometric luminosity, mass surface density of the clump environment and current protostellar mass (obtained from the average ``good'' model fits). The optimal (SOMA) aperture for each source, in arcseconds, is also included.}
    \begin{tabular}{lccccccc}
    \hline\hline
         Region & Source & $70{\rm \mu m}?$  & Opt. ap ($^{\prime\prime}$) & $\Sigma_{\rm cl}\:[{\rm g\: cm}^{-2}]$  & $L_{\rm bol}\: [L_{\odot}]$ & $M_{\rm env}\: [M_{\odot}]$ & $m_\star\: [M_{\odot}]$\\
    \hline
    AFGL 437 & A & Y & 16.75 & 0.67 & 4.205e+04 & 41.63 & 13.47\\
     & B & Y & 19.0 & 0.44 & 4.07e+03 & 37.22 & 4.18\\
    \hline
   AFGL 4029 &  & Y & 17.75 & 0.46 & 8.15e+04 & 140.69 & 19.74\\
    \hline
    AFGL 5180 & A & Y & 24.25 & 0.14 & 3.87e+04 & 152.14 & 15.92\\
    & B & Y & 14.25 & 0.58 & 1.02e+04 & 32.64 & 6.60\\
    \hline
    G18.67 & A & Y & 18.25 & 0.38 & 1.10e+05 & 85.03 & 23.30\\
    & B & Y & 17.25 & 0.53 & 2.13e+05 & 143.10 & 30.89\\
    & C & Y & 12.75 & 0.80 & 1.70e+05 & 156.79 & 25.95\\ 
    & D & Y & 17.5 & 1.058 & 3.22e+05 & 225.22 & 33.93\\
    & E & Y & 11.0 & 0.74 & 1.14e+05 & 139.49 & 21.65\\
    \hline
    G28.20 & A & Y & 15.5 & 0.76 & 4.31e+05 & 180.47 & 43.68\\
     & B & Y & 16.5 & 0.62 & 3.36e+04 & 81.56 & 11.78\\
     & C & Y & 16.0 & 0.45 & 8.94e+03 & 63.05 &  6.21\\
     & D & Y & 18.25 & 0.66 & 6.34e+04 & 99.64 & 16.25\\
     & E$^{**}$ & N & 15.5 & - & - & - & - \\
    \hline
    G30.76 & A & Y & 24.0 & 0.38 & 9.66e+04 & 72.74 & 21.99\\
    & B & Y & 16.0 & 0.63 & 8.14e+04 & 118.22 & 18.78\\
    & C & Y & 13.5 & 0.60 & 9.21e+04 & 111.29 & 20.26\\
    \hline
    G32.03 & A & Y & 12.5 & 0.41 & 1.59e+05 & 125.32 & 27.44\\
    & B & Y & 18.5 & 0.46 & 7.68e+03 & 60.24 & 5.72 \\
    & C & Y & 18.25 &  1.03 & 2.01e+05 & 251.25 & 26.64\\
    & D & Y & 17.25 & 0.41 & 5.05e+03 & 52.03 & 4.69\\
    \hline
    G35.03 &  & Y & 15.0 & 0.25 & 8.49e+04 & 196.62 & 20.22\\
    \hline
   G40.62 &  & Y & 13.25 & 1.22 & 5.31e+05 & 212.49 & 44.06 \\
    \hline
   G49.27 & & Y & 17.25 &  0.45 & 1.34e+05 & 100.16 & 25.22\\
    \hline
    G58.77 & A & Y & 21.5 & 0.14 & 8.57e+04 & 152.62 & 23.69\\
    & B & Y & 12.75 &  0.54 & 9.72e+03 & 34.55 & 6.47\\
    \hline
    IRAS 20343 &  & Y & 15.5 & 0.15 & 1.50e+04 & 20.17 & 10.46\\
    \hline
    S235$^*$ & & & 12.0 & 0.6 & 2.8e+04 & 6 & 12.4\\
    \hline
    \end{tabular}
    \tablefoot{$^*$S235 parameters are from \cite{liu2020sofia}. $^{**}$G28.2 E is not visible in the 70 $\rm \mu m$ data, thus the used optimal aperture is the same as for G28.2~A.}
    \label{tab:SED_params}
\end{table*}


\section{Results}\label{sec:results_analysis}

\subsection{Mean magnetic field orientation and polarization fraction}

The mean magnetic field orientation within the SOMA aperture (see Table \ref{tab:SED_params}) was calculated after applying a signal-to-noise ratio (SNR) cut on intensity (SNR$_{I}$) and on polarization fraction (SNR$_P$). All further calculations are done with a SNR$_{I}>10$. For sources with more than four data points having SNR$_P >3$, the SNR$_P$ cut was set to SNR$_{P}> 3$. There are four sources with fewer than four data points meeting this threshold; for these we use a less stringent cut of SNR$_{P}> 2$. These sources are denoted with an asterisk ($^*$) in Table \ref{tab:basicprops}. We calculate the mean magnetic field orientation via \citep{panopoulou2021revisiting}: 
\begin{equation}
    \Bar{\theta}_w = \frac{1}{2}\arctan\left(\frac{\sum_{i=1}^N w_i\sin{2\theta_i}}{\sum_{i=1}^N w_i}, \frac{\sum_{i=1}^N w_i \cos{2\theta_i}}{\sum_{i=1}^N w_i}\right),
    \label{eq:weight_B_angle_mean}
\end{equation}
where $N$ is the number of vectors inside the SOMA aperture and $w_i$ is the weight. For the magnetic field within the SOMA aperture we calculated the equally-weighted mean with $w_i=1$, as well as the intensity weighted mean with $w_i=I_i/I_{\rm max}$. Equally-weighted mean magnetic field orientations are shown in parenthesis in Table \ref{tab:basicprops}. The uncertainty in the mean orientation is then defined by propagation of the errors in the angle measurements, i.e.,
\begin{equation}
    \Delta\Bar{\theta}_w = \frac{1}{\sum_{i=1}^N w_i} \sqrt{\sum_{i=1}^N (w_i \cdot \sigma_{\theta_i})^2}.
    \label{eq:error_mean_weighted}
\end{equation}

The mean debiased polarization fractions and the corresponding error are similarly calculated with 
\begin{align}
    \Bar{p}_w & = \frac{1}{\sum_{i=1}^N w_i} \sum_{i=1}^N w_i p_i
    \label{eq:polfrac_averages}
\end{align}
and
\begin{align}
    \Delta\Bar{p}_w&= \frac{1}{\sum_{i=1}^N w_i} \sqrt{\sum_{i=1}^N (w_i\sigma_{p_i})^2 }.
    \label{eq:polfrac_averages_errors}
\end{align}

\subsection{Angular dispersion}

Angular dispersion, $D$, of the magnetic field orientation quantifies its variation. Here we use the dispersion function as introduced in \citet{2015A&A...576A.104P, polfracvdisp}:
\begin{equation}
    D(d) = \sqrt{\frac{1}{N} \sum_{i=1}^N [\theta(x+d_i)-\theta(x))]^2},
    \label{eq:dispersion}
\end{equation}
where $d$ is the distance from a given central position $x$. The dispersion can thus be seen as the difference in magnetic field orientation between two points separated by a distance $d$. Here we compute the angular dispersion using an aperture-based approach, and are thus limited by the beam size. As demonstrated in Appendix A, the grid size of the map has no significant effect and therefore we applied the following method to calculate the dispersion in a systematic way.

\begin{figure}
    \centering
    \includegraphics[width=\columnwidth]{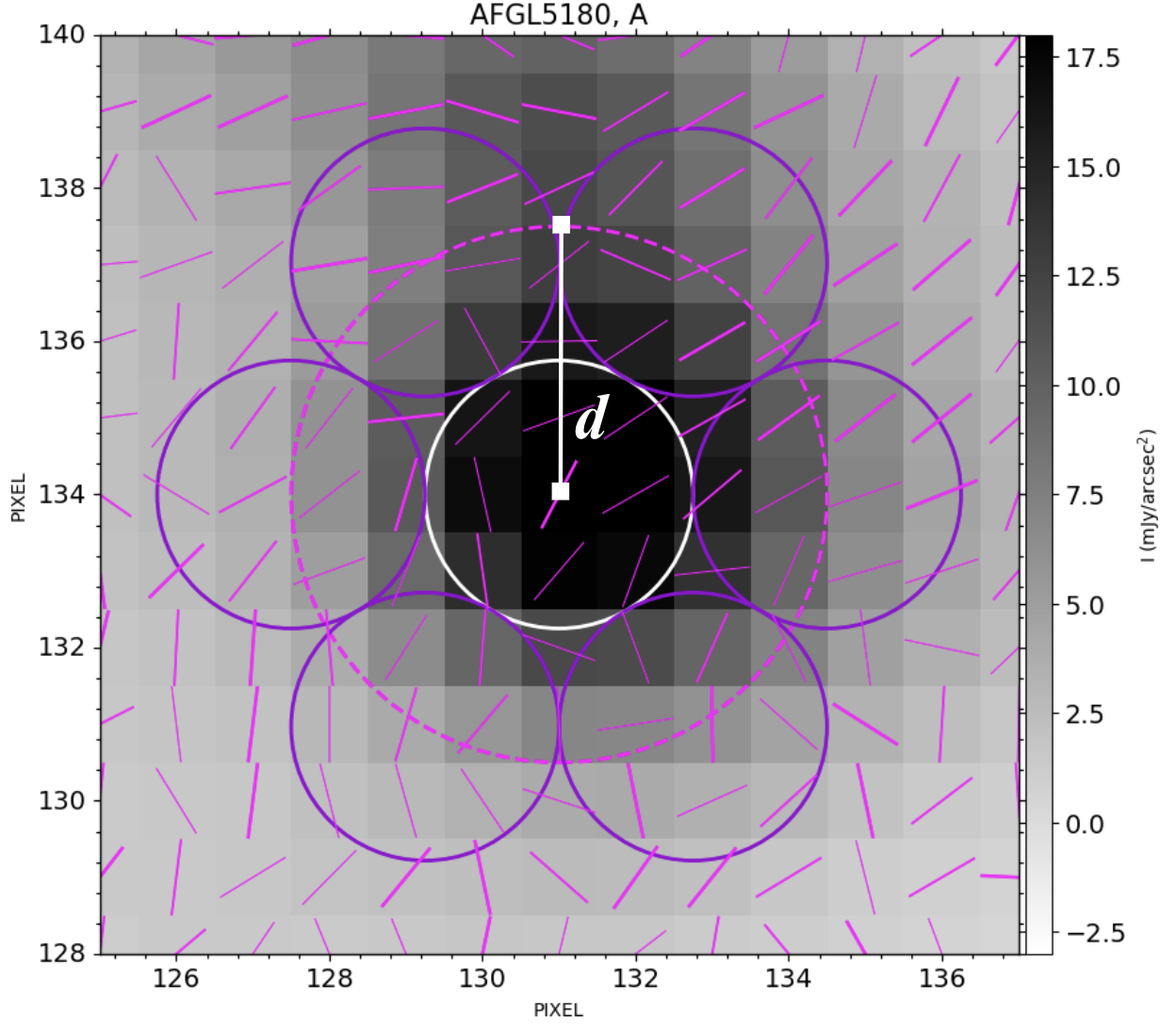}
    \caption{Illustration of dispersion calculations using AFGL~5180 A as an example, for a distance $d$, where $d$ is twice the beam aperture. The magenta lines represent magnetic field orientations, with thicker lines having greater SNR. All vectors have SNR$_I>10$. The thickest lines have SNR$_P>3$, the medium thickness lines have $3\geq {\rm SNR}_P>2$, and the thinnest have SNR$_P\leq2$. The white central circle is the beam-sized aperture within which the mean magnetic field is calculated as the reference ($\theta(x)$ in eq.~\eqref{eq:dispersion}). The mean magnetic field inside the surrounding purple beam-sized circles represents $\theta(x+d_i)$ in eq.~\eqref{eq:dispersion}. }
    \label{fig:disp_ill}
\end{figure}

First, a beam-sized aperture (with radius $r=7^{\prime\prime}$) is centered at the source intensity peak. The mean magnetic field orientation is calculated within this aperture using eq.~\eqref{eq:weight_B_angle_mean} with $w_i=1$. No ${\rm SNR}_P$ cut is made here, due to the small sample on vectors within each aperture. We define this as the source mean vector, corresponding to $\theta(x)$ in eq.~\eqref{eq:dispersion}. The next step is to obtain the magnetic field orientation at a distance $d$ from this central point. For this purpose we defined a larger circle with radius $d$ centered at the same position as the beam-sized aperture. We then place smaller, beam-sized apertures, on the circumference of this larger circle. The number of beam-sized circles is determined by the distance $d$. First, we fit as many full circles on the circumference of the large circle as possible without overlap. This number $N'$ is obtained via:
\begin{equation}
    N' = (2\pi d) // (2 r),
    \label{eq:nr circles}
\end{equation}
where $//$ denotes integer division. $N'$ is now the maximum number of beam-sized apertures that can fit on the circumference of the larger circle without overlapping. Now, if the remainder of the division is greater than half a beam size, i.e., if 
\begin{equation*}
    (2\pi d) \bmod (2r) \geq 8,
\end{equation*}
we add one more circle to the circumference, tolerating the small amount of overlap. In the end we have $N$ beam-sized apertures surrounding the central aperture. See Figure~\ref{fig:disp_ill} for an illustration of this procedure. The mean magnetic field orientation within every surrounding circle is then calculated with eq.~\eqref{eq:weight_B_angle_mean}, with $w_i = 1$. These values corresponds to $\theta(x+d_i)$ in eq.~\eqref{eq:dispersion}, which is then used to obtain the dispersion for the distance $d$. 

The uncertainty $\sigma_D$ in the dispersion is derived following \cite{alina2016polarization}. We first define $\Delta \theta(x+d_i)$ as the error in $\theta(x+d_i)$ and $\Delta \theta(x)$ as the error in $\theta(x)$, calculated with eq.~\eqref{eq:error_mean_weighted}.
The uncertainty in the dispersion is then given by
\begin{align}
    \sigma_D &= \frac{1}{N \cdot D} \left[ \left( \sum_{i=1}^N (\theta(x) - \theta(x + d_i)) \right)^2 \Delta \theta(x)^2 \right. \nonumber \\ 
    & \left. \quad +  \sum_{i=1}^N (\theta(x) - \theta(x + d_i))^2 \Delta \theta(x + d_i)^2  \right]^{1/2}
    \label{eq:error_disp}
\end{align}
Figure \ref{fig:disp_ill} illustrates the method used for dispersion calculation for source AFGL 5180 A. The dispersion was calculated for two values of $d$, i.e., equal to the SOMA aperture and two times the beam aperture ($14"$). These results are listed as $D_{\rm Beam}$, and $D_{\rm SOMA}$ in Table \ref{tab:basicprops}.

\subsection{Magnetic field and polarization properties}
Table \ref{tab:basicprops} shows the calculated basic properties for each source. These include average values of the Stokes parameters, the mean magnetic field orientation and dispersion, as well as the average biased and debiased polarization fraction. All calculations were performed within the SOMA apertures (see Table \ref{tab:SED_params}), and the dispersion calculation was also done for the beam aperture. Values for mean magnetic field orientation and polarization fraction are intensity weighted, with values in parentheses being equally weighted.

\begin{sidewaystable*}
\setlength{\tabcolsep}{1.5pt}
\centering
\caption{Summary table of basic properties for all sources}
\begin{tabular}{lcccccccccccccc}
    \hline\hline
    Region & Source & $I$ & $\Delta I$ & $Q$ & $\Delta Q$ & $U$ & $\Delta U$ & $\bar{\theta}$ & $\Delta\bar{\theta}$ & $D_{\rm Beam}$ & $D_{\rm SOMA}$ & $\bar{p}$ & $\Delta \bar{p}$ & $\bar{p}'$\\ 
     & & (mJy as$^{-2}$) & (mJy as$^{-2}$) & (mJy as$^{-2}$) & (mJy as$^{-2}$) & (mJy as$^{-2}$)& (mJy as$^{-2}$) & (deg) & (deg) & (deg) & (deg) & (\%) & (\%) & (\%)\\
    \hline
    AFGL 437 & A & 3.5341 & 0.0052 & 0.0028 & 0.006 & 0.0133 & 0.0056 & -61.0 (-66.5) & 3.3 (3.3) & 63.61 $\pm$ 4.07 & 52.01 $\pm$ 4.57 & 4.5 (4.8) & 0.4 (0.5) & 4.3 (4.6)\\
    & B & 2.5061 & 0.0048 & -0.0101 & 0.0053 & 0.0364 & 0.0049 & -42.1 (-35.7) & 2.4 (2.3) & 17.56 $\pm$ 7.28 & 30.65 $\pm$ 7.25 & 6.2 (6.9) & 0.4 (0.5) & 6.0 (6.6)\\
    \hline
   AFGL 4029 &  & 5.2268 & 0.0061 & 0.082 & 0.0063 & -0.0243 & 0.0058 & 79.3 (81.1) & 1.6 (1.5) & 28.83 $\pm$ 3.37 & 37.89 $\pm$ 3.04 & 3.1 (3.5) & 0.2 (0.2) & 3.0 (3.4)\\
    \hline
    AFGL 5180 & A & 7.3068 & 0.0059 & 0.0124 & 0.005 & 0.0707 & 0.0055 & -56.0 (-48.9) & 1.3 (1.2) & 29.51 $\pm$ 2.69 & 31.68 $\pm$ 2.11 & 3.7 (5.1) & 0.2 (0.2) & 3.6 (4.9)\\
    & B  & 3.9113 & 0.02 & -0.0017 & 0.02 & -0.0546 & 0.0222 & -12.2 (-20.5) & 3.5 (3.3) & 53.4 $\pm$ 4.51 & 53.4 $\pm$ 4.51 & 9.9 (10.7) & 1.0 (1.1) & 9.5 (10.3)\\
    \hline
    G18.67 & A & 3.0534 & 0.0057 & -0.0085 & 0.0051 & -0.0284 & 0.0053 & 27.8 (34.7) & 2.9 (2.6) & 66.15 $\pm$ 5.78 & 56.34 $\pm$ 3.74 & 4.2 (5.8) & 0.4 (0.7) & 4.0 (5.5)\\
    & B$^*$ & 2.1172 & 0.0064 & -0.0021 & 0.0058 & 0.0012 & 0.006 & 9.60 (13.9) & 3.6 (3.4) & 54.76 $\pm$ 4.86 & 49.21 $\pm$ 4.55 & 5.0 (6.6) & 0.6 (1.0) & 4.6 (6.0)\\
    & C & 1.9208 & 0.0135 & -0.0196 & 0.0117 & 0.0042 & 0.0121 & -16.4 (-17.8) & 3.5 (3.4) & 69.79 $\pm$ 4.54 & 58.49 $\pm$ 5.15 & 11.5 (13.0) & 1.5 (1.8) & 11.0 (12.4)\\ 
    & D & 3.3264 & 0.0188 & 0.0418 & 0.0151 & -0.0115 & 0.0157 & -55.8 (-73.6) & 5.0 (4.0) & 70.85 $\pm$ 4.06 & 62.05 $\pm$ 3.75 & 16.4 (21.9) & 2.4 (3.3) & 15.7 (21.0)\\
    & E$^*$ & 2.1739 & 0.0297 & -0.0242 & 0.0271 & -0.0761 & 0.0279 & 38.9 (36.0) & 5.1 (4.8) & 26.59 $\pm$ 13.3 & 41.09 $\pm$ 9.72 & 13.4 (15.1) & 2.6 (3.0) & 12.1 (13.6)\\
    \hline
    G28.20 & A & 16.0924 & 0.0136 & -0.0015 & 0.0073 & -0.1434 & 0.0066 & 40.9 (38.3) & 1.9 (1.9) & 24.99 $\pm$ 2.56 & 24.03 $\pm$ 2.15 & 1.8 (2.2) & 0.2 (0.3) & 1.7 (2.1)\\
     & B & 2.9118 & 0.0075 & 0.0305 & 0.0084 & -0.0171 & 0.0077 & -84.7 (34.9) & 4.6 (3.8) & 56.93 $\pm$ 4.41 & 52.93 $\pm$ 4.92 & 8.2 (12.7) & 1.0 (1.6) & 7.9 (12.2)\\
     & C & 1.8575 & 0.0115 & -0.0471 & 0.0129 & 0.0119 & 0.0116 & -6.5 (-5.3) & 3.4 (3.3) & 30.27 $\pm$ 4.59 & 43.87 $\pm$ 5.02 & 14.1 (16.5) & 1.5 (2.0) & 13.5 (15.8)\\
     & D & 1.619 & 0.0078 & 0.0036 & 0.0088 & -0.0053 & 0.008 & 51.4 (45.0) & 2.7 (2.7) & 61.25 $\pm$ 4.89 & 54.05 $\pm$ 6.58 & 16.6 (17.7) & 1.3 (1.4) & 16.1 (17.2)\\
     & E & 1.5683 & 0.0084 & 0.0362 & 0.0088 & -0.0164 & 0.008 & 79.9 (79.0) & 2.6 (2.5) & 51.96 $\pm$ 5.49 & 49.04 $\pm$ 4.45 & 15.3 (16.7) & 1.3 (1.5) & 14.8 (16.1)\\
    \hline
    G30.76 & A & 2.0756 & 0.0042 & 0.0179 & 0.0045 & -0.0226 & 0.0042 & 73.2 (71.2) & 2.2 (2.0) & 31.65 $\pm$ 4.49 & 38.04 $\pm$ 5.34 & 7.2 (9.6) & 0.5 (0.7) & 7.0 (9.3)\\
    & B$^*$ & 6.0672 & 0.0097 & -0.0279 & 0.0077 & -0.0215 & 0.0073 & 21.1 (13.8) & 2.4 (2.4) & 28.05 $\pm$ 4.9 & 28.39 $\pm$ 3.16 & 2.6 (3.1) & 0.3 (0.4) & 2.4 (2.9)\\
    & C & 1.9348 & 0.0099 & 0.0162 & 0.0115 & -0.0131 & 0.0108 & -82.7 (87.5) & 2.5 (2.3) &  63.0 $\pm$ 3.69 & 59.12 $\pm$ 3.29 & 11.3 (13.0) & 0.8 (0.9) & 10.9 (12.5) \\ 
    \hline
    G32.03 & A$^*$ & 13.2514 & 0.0186 & 0.017 & 0.0117 & -0.0547 & 0.0108 & 61.7 (67.1) & 2.9 (2.9) & 39.11 $\pm$ 8.77 & 49.7 $\pm$ 3.94 & 2.2 (2.4) & 0.3 (0.4) & 2.1 (2.2)\\
    & B & 2.9395 & 0.0071 & 0.0704 & 0.0076 & -0.0069 & 0.007 & 79.9 (80.7) & 2.1 (2.1) & 49.72 $\pm$ 10.98 & 23.83 $\pm$ 10.94 & 7.3 (7.7) & 0.5 (0.5) & 7.0 (7.5)\\
    & C & 3.3615 & 0.0155 & 0.0211 & 0.0161 & 0.0037 & 0.0144 & 37.7 (37.8) & 3.0 (3.0) & 62.74 $\pm$ 6.53 & 56.57 $\pm$ 5.31 & 15.5 (20.2) & 1.6 (2.2) & 14.9 (19.4)\\
    & D & 1.1139 & 0.0109 & 0.0156 & 0.0121 & 0.0173 & 0.0112 & -8.0 (0.0) & 3.3 (3.5) & 47.05 $\pm$ 5.8 & 47.61 $\pm$ 9.93 & 27.2 (28.0) & 2.7 (3.1) & 26.5 (27.2)\\
    \hline
    G35.03$^*$ &  & 7.8558 & 0.0108 & 0.0034 & 0.0073 & -0.0427 & 0.0067 & 51.6 (51.5) & 2.9 (2.7) & 31.14 $\pm$ 8.12 & 38.5 $\pm$ 8.25 & 2.1 (3.2) & 0.3 (0.5) & 1.9 (2.9)\\
    \hline
   G40.62 &  & 5.6899 & 0.0086 & -0.0325 & 0.0107 & -0.0044 & 0.0098 & 11.9 (14.4) & 2.7 (2.7) & 43.97 $\pm$ 3.63 & 41.57 $\pm$ 3.09 & 3.4 (3.7) & 0.3 (0.3) & 3.2 (3.6)\\
    \hline
   G49.27 &  & 7.4423 & 0.0063 & 0.1732 & 0.006 & -0.156 & 0.0059 & 69.1 (70.5) & 0.7 (0.8) & 11.33 $\pm$ 1.48 & 10.24 $\pm$ 1.83 & 3.6 (3.8) & 0.1 (0.1) & 3.5 (3.8)\\
    \hline
    G58.77 & A & 2.1823 & 0.0044 & 0.0021 & 0.0046 & -0.006 & 0.0047 & 60.7 (41.5) & 2.3 (2.0) & 49.12 $\pm$ 3.87 & 68.85 $\pm$ 3.23 & 8.3 (11.5) & 0.5 (0.8) & 8.0 (11.1)\\
    & B & 1.1346 & 0.0089 & -0.0167 & 0.01 & 0.0002 & 0.0102 & -18.3 (-21.3) & 3.7 (3.6) & 49.76 $\pm$ 4.89 & 51.38 $\pm$ 4.21 & 15.9 (18.4) & 1.8 (2.2) & 15.2 (17.6)\\
    \hline
    IRAS 20343$^*$&  & 4.1523 & 0.0062 & -0.0089 & 0.0061 & -0.0344 & 0.0065 & 37.3 (39.0) & 2.7 (2.7) & 28.46 $\pm$ 13.17 & 32.08 $\pm$ 4.69 & 2.8 (2.9) & 0.2 (0.3) & 2.6 (2.7)\\
    \hline
    S235 & & 11.0202 & 0.0161 & 0.0677 & 0.0112 & 0.1035 & 0.0113 & -57.6 (-57.5) & 2.9 (2.8) & 29.5 $\pm$ 3.02 & 30.45 $\pm$ 4.63 & 2.5 (2.6) & 0.3 (0.3) & 2.4 (2.5)\\
    \hline
\end{tabular}
\tablefoot{Sources marked with $^*$ have a $P$ SNR cut $>2$ instead of $>3$. Values in parentheses are means evaluated with equal weight.}
\label{tab:basicprops}
\end{sidewaystable*}


\section{Analysis \& Discussion}\label{sec:analysis}
\begin{figure*}
    \includegraphics[width =\textwidth]{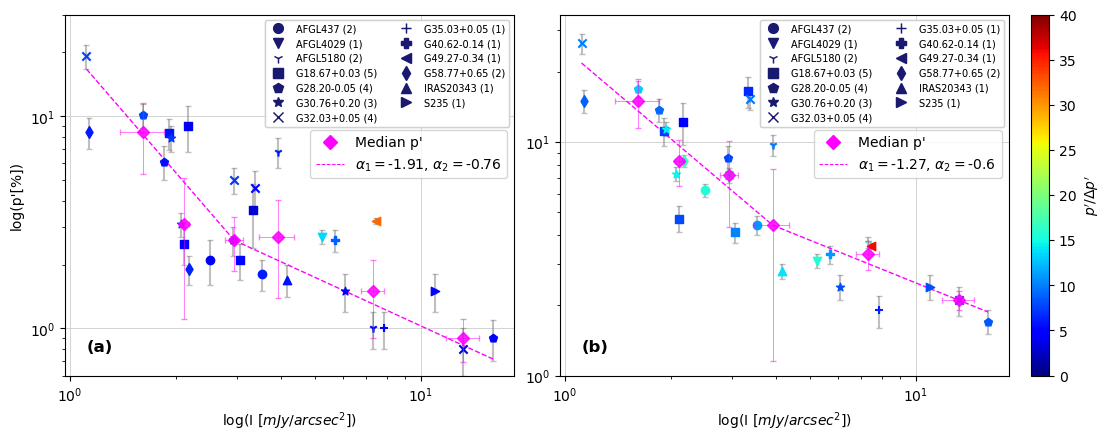}
    \caption{Average debiased polarization fraction within beam aperture ($a$) and SOMA aperture ($b$) plotted against Stokes intensity. The data points are color coded by signal to noise ratio in polarization fraction, and data points for sources in the same region are marked with the same symbols. A broken power law with exponents $\alpha_1, \alpha_2$ (see legend) was fitted to the running median of every five data points. The data points for the running medians and power law fit are shown in magenta.}
    \label{fig:p_v_I}
\end{figure*}

\subsection{Polarization fraction and intensity}

Dust grains are typically expected to align with their long axes perpendicular to the magnetic field, making it possible to infer the magnetic field orientation from polarized dust emissions \citep{davis1951polarization}. The dust grain alignment efficiency can be quantified by $p' \propto I^{-\alpha}$ \citep{andersson2015interstellar},  where the value of $\alpha$, expected to be in the range $0\leq\alpha \leq 1$, increases as dust grain alignment efficiency decreases. At low ${\rm SNR}_P$ we expect $p \propto 1/I$ (i.e., $\alpha \simeq 1$) from the Ricean noise distribution on $p$ \citep{pattle2019jcmt}. 

We present in Figure \ref{fig:p_v_I} the $p' \propto I^{-\alpha}$ plots of the SOMA regions at beam aperture (panel $a$) and SOMA aperture (panel $b$). Both plots are color-coded by the ${\rm SNR}_P$. We fit a broken power-law to the running median of every five data points for both $p' \propto I^{-\alpha}$ plots and we identified the break point manually by inspecting the point where the median $p'$ is clearly flattened. 

We find that in the low Stokes $I$ regime, where ${\rm SNR}_P$ is expected to be low, $\alpha >1$, suggesting that it is dominated by noise. However, in the high Stokes $I$ regime with good ${\rm SNR}_P$, the index is smaller, i.e., $\alpha = 0.76$ for beam aperture and $\alpha = 0.60$ for SOMA aperture, suggesting moderate grain alignment efficiencies. 

\subsection{Cloud-field alignment} 

\begin{figure*}
    \includegraphics[width=1 \linewidth]{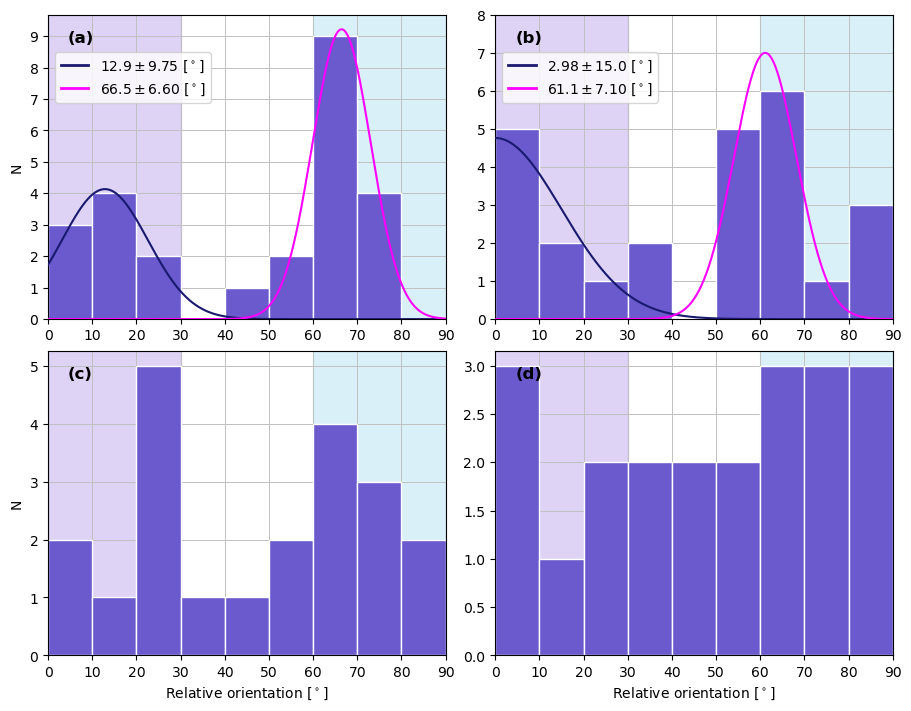}
    \caption{Sub-figures (a) and (b) shows the distribution of relative orientation between the magnetic field orientation and the source orientation derived using the Hessian matrix analysis, and the autocorrelation function respectively. Additionally two independent Gaussian were fitted to the distribution in sub-figures (a) and (b). The mean $\mu$, and standard deviation $\sigma$, for each fitted curve is indicated in the legend as $\mu \pm \sigma$. Similarly, figure 3c) and 3d) shows the distribution of the relative orientation between the magnetic field orientation and filament orientation, derived from Hessian analysis (c) and the autocorrelation function (d). The purple background range form $0^\circ-30^\circ$, indicating parallel alignment, while the blue range from $60^\circ-90^\circ$ corresponding to perpendicular alignment. All angles are normalized to be in the range $0^\circ-90^\circ$.}   \label{fig:histogram_source_filament_orientation}
\end{figure*}

\begin{figure*}
  \includegraphics[width=1 \linewidth]{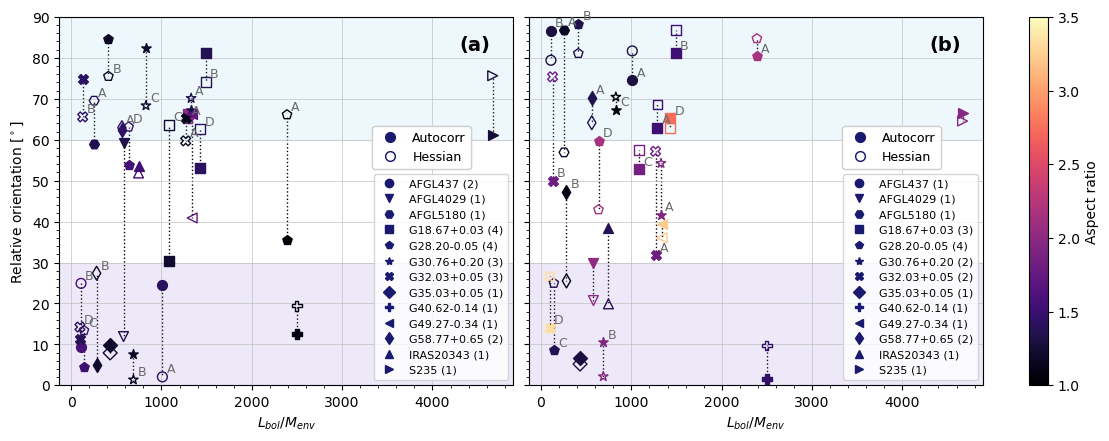}
  \caption{The figure shows the relative orientation between the source/filament orientation and average magnetic field orientation on the SOMA scale plotted against the ratio of bolometric luminosity to envelope mass. Figure (a) shows the relative source orientation, and figure (b) shows the relative filament orientations. The data points are color-coded by aspect ratio, with sources from the same regions represented by identical markers. The orientations obtained from the autocorrelation function are displayed with filled markers, while the orientations from the Hessian matrix analysis are represented with unfilled markers.}
  \label{fig:source_filament_orientation}
\end{figure*} 

Here we study the relative orientation between the mean magnetic field direction within the SOMA aperture and the elongation of the structure, as measured in both the SOMA aperture and the larger-scale filamentary structure. The structure elongation was estimated by two methods: Hessian matrix analysis \citep{soler2020history, Soler2013} and autocorrelation function \citep{li2013link}.

Filaments were identified in the Stokes $I$ maps using Hessian matrix analysis, as described by \citet{soler2020history}. In practice, we use the Hessian matrix analysis function in the AstroHOG package \citep{2019A&A...622A.166S}. Data points with SNR$_{I}<10$ were masked out to more accurately define the shapes of the filaments identified with the Hessian matrix analysis. By combining these two approaches, we were able to define the filament shapes and proceed with analyzing their orientation. The source boundary is simply defined by the optimal aperture obtained from the SED-analysis in \S\ref{sec:SED}. Only data points with SNR$_{I}<10$ are used here as well. The Hessian matrix analysis classifies each pixel as either being filamentary or not, and thus, in the process, assigns an orientation to each pixel. As a result, the orientation of the filament and the source can be obtained by calculating the average orientation using eq.~\eqref{eq:weight_B_angle_mean} within each of their respective boundaries. 

The orientations of the 
source and filaments were also calculated using the autocorrelation function of the Stokes $I$ map as described by \cite{li2013link}. First the Stokes $I$ map was Fourier transformed and multiplied by its complex conjugate. An inverse Fourier transform was then performed on the product in order to obtain the autocorrelation function. The result was then shifted for the filament to be centered. Finally, the autocorrelation was normalized by dividing by the number of pixels times the squared average intensity. The filament/source orientation is then given by the orientation of the long axis of the autocorrelation function, as defined in \cite{li2013link}. In order to obtain the orientation of the long axis, an ellipse was fitted to a contour of the autocorrelation function using the python package \textit{OpenCV}. This process was repeated for multiple contour levels, allowing estimation of the systematic error in the method by comparing the variation in orientation across different contour levels. The orientation of the ellipse then corresponds to the orientation of the filament or source. 

Table \ref{tab:orientations} presents the filament and source orientations derived from Hessian matrix analysis and the autocorrelation function of the Stokes $I$ map. The orientation from the autocorrelation function is calculated at three different contour levels, with the variations between them serving as an indication of error in the method. For sources located outside filaments identified by the Hessian matrix, no results are provided, and these cases are marked with a $-$. Blank rows indicate sources that lie within a filament already analyzed. Sources within the same region and filament are denoted by $^{\dagger}$. The relative orientation between the mean magnetic field orientation and the structure elongation within the SOMA aperture ($\theta_{rel,\rm SOMA}$) as well as filament elongation ($\theta_{rel,\rm Fil}$) are also summarized in Table~\ref{tab:orientations}. When using the average filament/source orientation obtained from the autocorrelation function, the average of the different contour levels is used. 

\setlength{\tabcolsep}{3pt}
\begin{table*}
\footnotesize
\centering
  \caption{Filament and source orientation obtained from Hessian Matrix (HM) analysis and the autocorrelation function of the Stokes $I$ map. For the autocorrelation function, the orientation was obtained using 3 different contour levels: 0.05, 0.1 and 0.2 for the filament orientation; and 0.1, 0.2 and 0.3 for the source orientation. The aspect ratio ($R_{\rm aspect}$) of the fitted ellipse is also shown, and values in parentheses indicate the contour level at which the aspect ratio of the source was determined. The relative orientation ($\theta_{\rm rel}$) between source/filament orientation and the average magnetic field orientation at the SOMA scale from Table \ref{tab:basicprops} is also displayed for the two methods separately.}
    \begin{tabular}{lcccccccccccccccc}
    \hline\hline
         Region & Source & \multicolumn{4}{c}{Filament Orientation} &\multicolumn{2}{c}{$\theta_{rel,\rm Fil}$} & & \multicolumn{5}{c}{Source Orientation} &\multicolumn{2}{c}{$\theta_{rel,\rm SOMA}$} \\
         \cline{3-9} \cline{11-17}
         & & HM $\mathbf{[^\circ]}$ & \multicolumn{3}{c}{Autocorrelation $\mathbf{[^\circ]}$} & $\theta_{rel,\rm HM}$ & $\theta_{rel,Auto}$ &  $R_{aspect}$ & & HM $\mathbf{[^\circ]}$ & \multicolumn{3}{c}{Autocorrelation $\mathbf{[^\circ]}$} &  $\theta_{rel,\rm HM}$ & $\theta_{rel,Auto}$ & $R_{aspect}$ \\
        \cline{4-6}  \cline{12-14}
        & & & 0.05 & 0.1 & 0.2 & & & & & &  0.1 & 0.2 & 0.3 \\
    \hline
    AFGL 437&  & 37.31 & 40.88 & 45.38 & 47.29 & 80.55 & 80.55 &1.30 & & - & - & -& - & - & - &- \\
     & A$^{\dagger}$ &  &  &  &  & 81.69 & 74.48 &  & & -63.07 & -87.44 & -81.61 & -87.70 & 2.07 & 24.59 & 1.41 (0.3) \\
     & B$^{\dagger}$ & & & & & 79.41 & 86.62 & & & -67.00 & -61.33 & -68.60 & 6.44 & 24.9 & 9.37 & 1.58 (0.4)\\
    \hline
   AFGL 4029 & & -79.99 & -74.69 & -67.28 & -70.53 & 20.71 & 29.87 & 2.01 & & -88.79 & 32.21 & 12.35 & 15.93 & 11.91 & 59.22 &  1.34 (0.3)\\
    \hline
    AFGL 5180 & A & 67.09 & 33.3 & 31.61 & 27.65 & 56.91 & 86.85 & 1.17 & & 13.54 & 2.20 & 3.70 & 3.32 & 69.54 & 59.07 & 1.36 (0.3) \\
    & B & - & - & - & - & - & - & - & & - &  -90.00 & -90.00 & -90.00 & - & 77.8 & 1.50 (0.3) \\
    \hline
    G18.67&  & -83.58 & -88.42 & -89.53 & -89.94 & 77.72 & 72.00 & 1.52 & &  &  & &  &  &  & \\
    & A$^{\dagger}$ &  &  & & & 68.62 & 62.90 & & &  -87.0 & -83.82 & - & -87.94 & 65.2 & 66.32 & 2.00 (0.3)\\
    & B$^{\dagger}$ & & & & & 86.82 & 81.10 & & & 83.80 & -87.44 & 90.00 & 90.00 & 74.2 & 81.25 & 1.37 (0.3) \\
    & C & 40.99 & 40.31 & 41.93 & 44.14 & 57.39 & 52.75 & 1.85 & & 47.22 & 0.00 & 0.00 & 48.87 & 63.62 & 30.39 & 1.22 (0.3) \\ 
    & D  & 7.03 & 8.04 & 9.53 & 10.91 & 62.83 & 65.29 & 2.72 & & 6.79 & 0.00 & -8.39 & -0.03 & 62.58 & 53.0 & 1.42 (0.4)\\
    & E  &  - & - & - & - & - & - & - & & - &  -65.35 & -79.49 & 90.0 & - & 62.81 & 1.28 (0.3)\\
    \hline
   G28.20 &  & -43.80 & -52.82 & -41.66 & -23.55 & 70.5 & 70.5 & 2.17 & &  &  & &  &  &  & \\
     & A$^{\dagger}$ &  & & & & 84.7 & 80.35 & & & -25.21 & 0.00 & 0.00 & 16.17 & 66.11 & 35.58 & 1.04 (0.3)  \\
     & B & -3.61 & 6.18 & 6.93 & 7.61 & 81.09 & 88.39 & 1.34 & & -9.21 & 0.00 & 0.00 & 0.00 & 75.49 & 84.7 & 1.16 (0.3)\\
     & C & -18.44 & -12.68 & -19.22 & -13.34 & 24.94 & 8.58 & 1.37 & & -19.88 & -16.62 & -8.39 & -7.86 & 13.38 & 4.45 & 1.52 (0.3) \\
     & D & -8.49 & -7.70 & -8.08 & -9.36 & 42.91 & 59.78 & 2.16 & & -11.73 & -5.28 & 0.06 & -2.06 & 63.13 & 53.83 & 1.55 (0.4)\\
     & E$^{\dagger}$ & & & & & 56.30 & 60.65 & & &  -59.92 & -67.00 & -75.36 & -82.19 & 40.18 & 25.24 & 1.29 (0.3)\\
     
    \hline
    G30.76&  & 18.95 & 38.10 & 36.64 & 19.70 & 28.2 & 26.05 & 1.93 & &  &  & &  &  &  & \\
     & A$^{\dagger}$ &  &  &  & & 54.25 & 41.64 & & & 3.07 & 5.76 & 6.28 & - & 70.13 & 67.18 & 1.33 (0.3) \\
    & B$^{\dagger}$ & & & & & 2.15 & 10.46 & & &  19.73 & 13.31 & 14.57 & 12.57 & 1.37 & 7.62 & 1.16 (0.3) \\
    
    & C & -12.27 & -18.39 & -16.53 & -11.33 & 70.43 & 67.28 &  1.03 & & -14.36 & 0.00 & 0.00 & -0.55 & 68.34 & 82.52 & 1.21 (0.4) \\ 
    \hline
    G32.03&  & 4.53 & 31.24 & 29.58 & 29.08 & 65.27 & 40.83 & 1.82 &  &  & &  &  &  & \\
     & A$^{\dagger}$ &  &  &  &  & 57.17 & 31.73 & & & 1.98 & 0.00 & 0.00 & -10.51 & 59.72 & 65.19 & 1.14 (0.3)\\
    & B$^{\dagger}$ & & & & & 75.37 & 49.93 & & &  14.32 & -31.12 & -19.83 & -24.62 & 65.58 & 74.91 & 1.42 (0.3)\\
    & C &  - & - & - & - & - & - & - & & - &  -45.0 & -47.79 & 82.76 & - & 81.6 & 1.27 (0.3)\\
    & D &  18.60 & 2.73 & 4.10 & 11.23 & 26.6 & 14.01 & 3.33 & & 6.24 & 2.56 & 7.54 & 0.03 & 14.24 & 11.37 & 1.39 (0.3)\\
    \hline
    G35.03 & & 46.40 & 45.00 & 45.00 & 45.00 & 5.2 & 6.60 &  1.30 & & 43.68 & 45.0 & 35.13 & 45.0 & 7.92 & 9.88 & 1.21 (0.1)\\
    \hline
   G40.62 & & 21.64 & 16.00 & 9.18 & 14.97 & 9.74 & 1.49 & 1.44 & & 31.32 & 0.0 & 0.0 & -89.45 & 19.42 & 12.45 & 1.07 (0.2)\\
    \hline
   G49.27 &  & 33.05 & 29.88 & 29.61 & 29.53 & 36.05 & 39.43 & 3.22 & & 28.21 & 0.0 & 8.39 & 0.03 & 40.89 & 66.3 & 1.71 (0.3)\\
    \hline
    G58.77 & A & -3.45 & -10.79 & -0.27 & -17.38 & 64.15 & 70.20 & 1.36 & & -2.24 &  0.0 & -5.29 & 0.0 & 62.94 & 62.46 & 1.43 (0.3)\\
    & B & 7.20 & -70.33 & -59.80 & -66.78 & 25.5 & 47.34 & 1.08 & & 9.07 & 0.0 & 0.0 & -45.0 & 27.39 & 5.02 & 1.21 (0.3)\\
    \hline
    IRAS 20343 &  & 57.20 & 74.93 & 75.98 & 76.02 & 19.9 & 38.34 & 1.37 & & 89.17 & 90.0 & -87.70 & 90.0 & 51.87 & 53.47 & 1.54 (0.3)\\
    \hline
    S235 & & 6.96 & 7.59 & 6.30 & 12.89 & 64.56 & 66.52 & 1.97 & & 18.03 & 0.0 & 0.0 & 10.51 & 75.63 & 61.09 & 1.26 (0.3)\\
    \hline
    \end{tabular}
    \tablefoot{Sources marked with $^{\dagger}$ are within the same filament structure. The corresponding relative orientations presented of each source within the same filament is defined by the mean magnetic field orientation of the individual source with respect to the filament orientation. Some sources lie outside filaments identified by the Hessian matrix analysis and thus lack values for filament orientation. Thus, there is also no source orientation from the Hessian matrix analysis for these sources. }
    \label{tab:orientations}
\end{table*}

The aspect ratio (ratio between major and minor axis) of filaments and sources is an important property when looking at orientation, as a circular source does not have a well defined orientation and can thus be affecting the results and complicating the interpretation. To obtain the aspect ratio, an ellipse was fitted to the intensity map of the filament and source respectively. For the filament orientation an ellipse was fitted directly to the filament identified with the Hessian matrix analysis and SNR$_{I}>10$, making it insensitive to contour level as all but the relevant pixels were masked out, and the value of all relevant data points were set to one. To obtain the aspect ratio for the sources, an ellipse was fitted to the intensity map within the SOMA aperture. The fitting process is sensitive to the chosen contour level, with a default value set to 0.3, as this provided good fits for most sources. In cases where the default level did not produce an accurate fit, the contour level was adjusted to better capture the shape of the source. The specific contour level used for each source is indicated in parentheses in Table \ref{tab:orientations}.

\subsubsection{Statistical tests}

The magnetic field is said to be aligned parallel with the source/filament if $\theta_{rel,\rm SOMA}$/$\theta_{rel,\rm Fil}$ is between $0-30^\circ$ and perpendicular if $\theta_{rel,\rm SOMA}$/$\theta_{rel,\rm Fil}$ is between $60-90^\circ$. Figure \ref{fig:source_filament_orientation} presents the distributions of $\theta_{rel,\rm SOMA}$ and $\theta_{rel,\rm Fil}$ with the structure orientations defined by the Hessian matrix analysis and autocorrelation, respectively. We tried and successfully fit two independent Gaussians to the histogram distributions of $\theta_{rel,\rm SOMA}$ (panels $a$ and $b$) as these show a strong tendency of bimodality. The fitted mean and dispersion of the two Gaussians are presented in legends in the two panels respectively.

We quantify the statistical significance of these bimodal distributions against a random uniform distribution via one-sample Kuiper test. The one-sample Kuiper test compares against the cumulative distribution function of a uniform distribution between values $[0,90]$, that is
\begin{equation}
    F(x) = \frac{x}{90}, 0\leq x \leq 90
    \label{cdf_uniform}
\end{equation}. 
The resulting $p$-value represents the probability of the data coming from the cumulative distribution function in \eqref{cdf_uniform}. The $\theta_{rel,\rm SOMA}$ distribution obtained from the Hessian matrix analysis has a $p$-value of $0.01$, and a $p$-value of $0.05$ from the autocorrelation function method. For $\theta_{rel,\rm Fil}$, the Hessian matrix method yielded a $p$-value of $0.45$, and a $p$-value of $0.81$ for the autocorrelation function method. 

We further applied a Monte Carlo simulation following similar methods as described in \citet{li2013link}. For each histogram distribution defined in Figure 3, we computed the observed ratio ($\theta_{rel,\rm obv}$) between the source number that have either relative orientations within 30 degrees from parallelism ($0^{\circ}-30^{\circ}$) or perpendicularity ($60^{\circ}-90^{\circ}$) and the remaining sources. We then randomly drew the number of vectors equal to the total source number, and constructed a histogram using the same bin width as in Figure \ref{fig:histogram_source_filament_orientation}. The ratio ($\theta_{rel,i}$) for this histogram was then computed and compared to $\theta_{rel,\rm obv}$. A $\theta_{rel,i}$ lower or equal to $\theta_{rel,\rm obv}$ indicates that the bimodal distribution results from a random uniform distribution at the same or more statistically significant than the observed bimodal distribution. We repeat the aforementioned step $10^{6}$ times to study the number of times $\theta_{rel,i}$ is smaller or equal to $\theta_{rel,\rm obv}$, which defines the statistical significance ($p-$value). For $\theta_{rel,\rm obv, SOMA}$, the $p-$values are $0.012$ and $0.045$ with the source elongation within the SOMA aperture estimated from the Hessian matrix analysis and from the autocorrelation function method respectively. For $\theta_{rel,\rm obv, Fil}$, the $p-$values are $0.12$ (Hessian matrix analysis), and $0.40$  (autocorrelation function). All the $p$-values are summarized in Table \ref{tab:orientation_pvalues}. The results are consistent between the two statistical tests, and we can independently from the methods used to derive the structural orientation, confirm a bimodal distribution for $\theta_{rel,\rm SOMA}$ but not for $\theta_{rel,\rm Fil}$. 

\begin{table}[h!]
    \centering
    \caption{$p$-values of the K-S/Monte-Carlo  tests performed in this work to compare the observed $\theta_{rel,\rm SOMA}$ and $\theta_{rel,\rm Fil}$ histogram distributions to a random uniform distribution.}
    \begin{tabular}{lccc}
        \hline\hline
         & Hessian & Autocorr \\
        \hline
        $\theta_{rel,\rm SOMA}$ & 0.010/0.012 & 0.050/0.045  \\
        \hline
        $\theta_{rel,\rm Fil}$ & 0.45/0.12 & 0.81/0.40 \\    
        \hline
    \end{tabular}
    \label{tab:orientation_pvalues}
\end{table}

Figure \ref{fig:source_filament_orientation} presents the relative orientation of each source inferred from the two methods (connected with a dotted line) plotted against the protostellar bolometric luminosity to envelope mass ratio, which relate to the protostellar evolutionary stages. No clear correlations have been identified in either the source or filament cases. We notice that while we confirm a bimodal orientation independent of the method to measure the structure orientations, we also observe significant changes in the relative orientation between the two methods for some sources. Hence, these findings further stress the needs to cross-check the structure orientation with multiple methods. 

The bimodal distribution of relative orientation between the SOMA source orientation and the mean local magnetic field orientation indicates that the magnetic field's regulation of source formation and evolution as proposed in \cite{li2013link} can be extended to high-mass star-forming regions. This suggests that $B$-fields regulate in similar ways in both high- and low-mass star forming molecular clouds. We note that one possible explanation for the lack of a clear bimodal orientation on the filament scale is that the POL-2 observations are not sensitive to the more diffuse low intensity regions. The observed bimodality at the SOMA source scale may be inherited down to smaller scales, where there is evidence that magnetic fields to continue to play a regulative role \citep[e.g.,][]{2014ApJ...792..116Z}.

\subsection{Inter-source angular difference in magnetic field orientation}

To study the magnetic field on larger scales the angular difference, $\Delta \theta$, of the magnetic field between individual pairs of neighboring sources within the same regions was calculated. That is the difference in magnetic field orientation between all individual pairs of sources within each region, as a function of the physical separations between the sources. The separation between each source pair was determined based on the pixel grid size of the map. The pixel separation between two sources was calculated with the Pythagorean theorem, and this was then converted to the separation in arcseconds (1 pix = $4^{\prime\prime}$). The physical separation was then converted to parsec using the distance $d$ (in parsec) from Table \ref{tab:sourceinfo} of the region. Note that here we assume each individual source within a region is at the same distance as the main SOMA source. The distance $\alpha$ in arcseconds between two sources was converted to radians $\alpha_{\rm rad}$ and used to convert this distance to parsec. For small angles $\sin(\alpha_{\rm rad})$ can be approximated with $\alpha_{\rm rad}$ and the distance in parsec between the sources is given by $d \cdot \alpha_{\rm rad}$.
\begin{figure}
  \includegraphics[width=1 \linewidth]{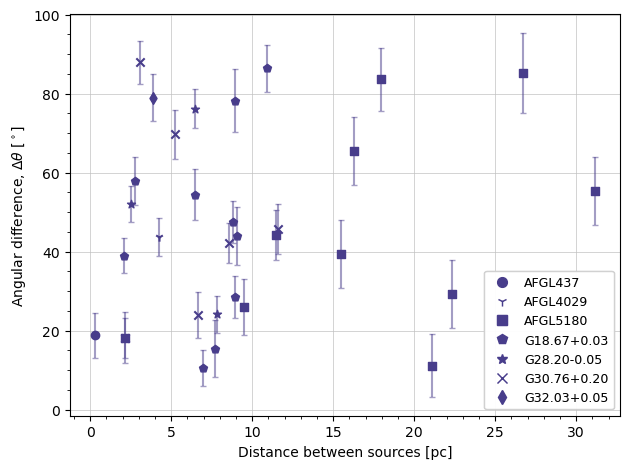}
  \caption{Angular difference between the magnetic field orientation for individual pair of sources within the same region as a function of the distance between the sources. The region of the pair is indicated by the marker.}
  \label{fig:rel_orientation_between_sources}
\end{figure}

We applied a Kuiper test to the distribution of angular difference comparing to a uniform distribution (eq.~\eqref{cdf_uniform}). The Kuiper statistic for the distribution of angular difference is $0.19$ with the corresponding $p$-value $0.62$, showing no clear correlations in the angular difference between sources. 

Our result differs from that of some previous studies of star-forming regions \citep{li2014link}. This may indicate that the source separations probe a scale at which gravitationally-induced motions controlling orbits in the protocluster potential begin to dominate over large scale magnetic field dynamics.

\subsection{Correlations between angular dispersion and other polarimetric metrics}

\begin{figure*}
  \includegraphics[width=1 \linewidth]{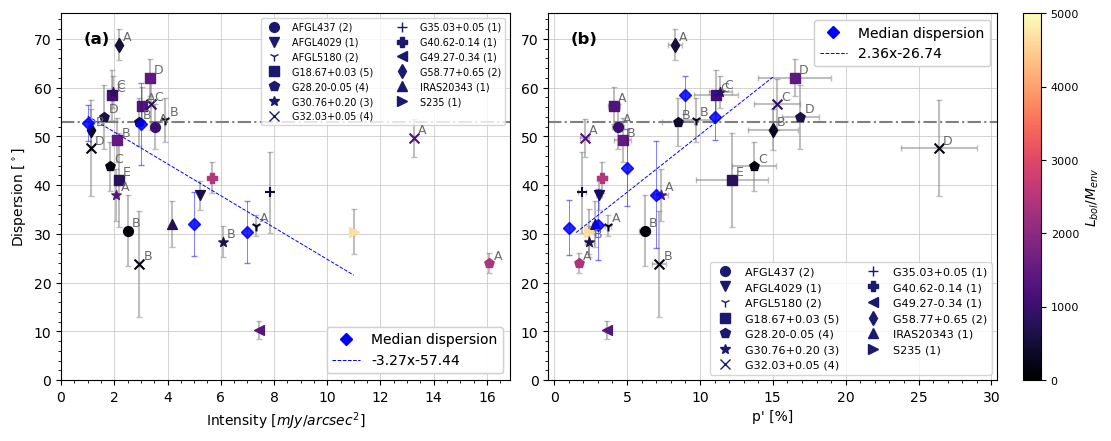}
  \caption{Angular dispersion at SOMA aperture plotted as a function of the mean intensity within the SOMA aperture (a) and debiased polarization fraction on SOMA scale (b). The data points are color-coded by $L_{\rm bol}/M_{\rm env}$ from Table \ref{tab:SED_params}. A line was fitted to the median dispersion within bins of size 2 (blue data points). The gray line represents the theoretical maximum dispersion.}
  \label{fig:disp_v_I_p}
\end{figure*}

In Figure \ref{fig:disp_v_I_p} we present 
plots of the angular dispersion as a function of Stokes $I$ intensity (see Table \ref{tab:basicprops}) and as a function of the debiased polarization fraction. The data points are color-coded by the bolometric luminosity to envelope mass ratio ($L_{\rm bol}/M_{\rm env}$) of each source. $L_{\rm bol}/M_{\rm env}$ represents an estimation of the evolutionary stage of the protostar \citep{zhang2018radiation}. We also compute a running median dispersion, $\Tilde{D}$, for both plots using a bin width of $p' = 2\%$, which is increased when necessary to ensure that each bin contains at least five data points. The median absolute deviation, MAD, was calculated using 
\begin{equation}
  {\rm MAD} = 0.67449 \sigma_{\Bar{D}},
    \label{std_error_median}
\end{equation}
where $\sigma_{\Bar{D}}$ is the standard deviation in the angular dispersion. We applied a least square fit (\textit{numpy.polyfit}) to the running median, and the slope of the fit is $-3.27$ and $2.02$ for the dispersion versus intensity and polarization fraction respectively (Figure \ref{fig:disp_v_I_p}).

The gray line in Figure \ref{fig:disp_v_I_p} represents the theoretical maximum dispersion for a random set of polarization angles \citep{serkowski1962polarization}. The deviation seen from this in Figure \ref{fig:disp_v_I_p} is likely due to the small sample of angles, following the Central Limit Theorem. 

The correlation between angular dispersion and Stokes $I$ intensity was also evaluated using a Spearman Rank test. This analysis yielded a correlation coefficient of $-0.48$ and a corresponding $p$-value of $0.0097$. This is the $p$-value corresponding to the null hypothesis that two samples have no correlation. These results suggest a statistically significant negative correlation between angular dispersion and intensity.

Similarly, the Spearman Rank test for the correlation between angular dispersion and the debiased polarization fraction gives a correlation coefficient of $0.58$ with a $p$-value of $0.001$, suggesting a statistically significant positive correlation. Admitting the large error bars, the overall trend visually differs from what has previously been seen in low-mass star-forming regions by \cite{ade2015planckXX}, which showed an inverse relationship between $p$ and $D$ for polarized emission in interstellar clouds. This apparently opposite trend does not necessarily contradict what was observed with {\it Planck}, as these trends presented in Figure \ref{fig:disp_v_I_p} are likely the result of the more noisy data toward regions with lower intensity, where also the polarization fractions are higher.

\subsection{Correlations between polarization and protostellar properties}

\begin{figure*}
    \centering
    \includegraphics[width=1\linewidth]{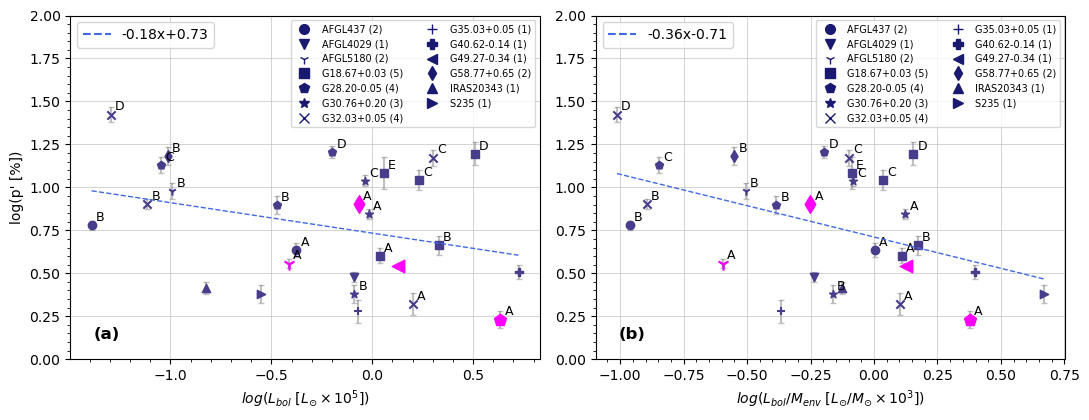}
    \caption{Polarization fraction on SOMA scale against the bolometric luminosity ($a$) and the ratio between bolometric luminosity and envelope mass ($b$) from Table \ref{tab:SED_params}. The dotted lines show fits to the data. The letter next to the data points specifies the source in the region given by the marker that the point refers to. Furthermore the four sources we have outflow directions for are plotted with larger, magenta symbols.}
    \label{fig:p_v_L_M}
\end{figure*}

\begin{figure*}
    \centering
    \includegraphics[width = 1 \textwidth]{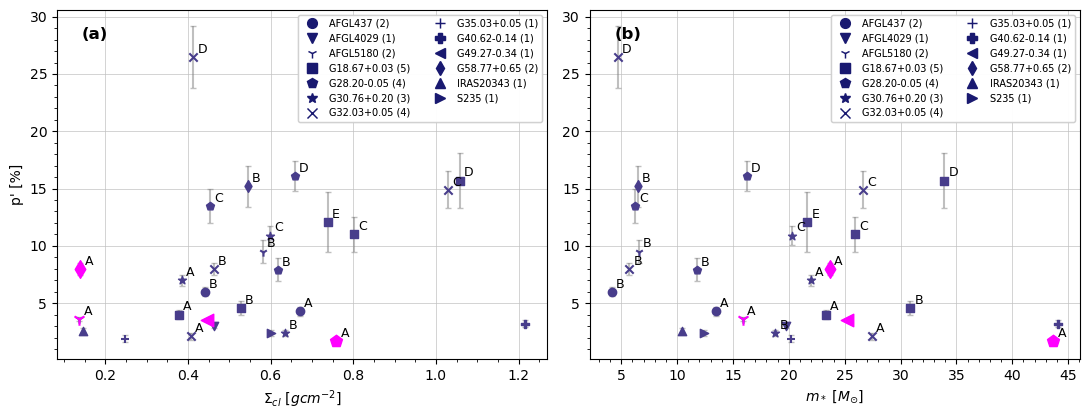}
    \caption{Polarization fraction on the SOMA scale against the surface density of the clump environment and star mass from Table \ref{tab:SED_params}. The four sources we have outflow direction for are plotted with larger, magenta symbols.}
    \label{fig:p_v_sigma_mstar}
\end{figure*}

In Figure \ref{fig:p_v_L_M} we present the plots of $p'$ as a function of bolometric luminosity and $L_{\rm bol}/M_{\rm env}$. A Spearman rank test for $p'$ v $L_{\rm bol}$ yielded a correlation coefficient of $-0.20$ with a $p$-value of $0.31$. The same test result for $p'$ vs $L_{\rm bol}/M_{\rm env}$ produced a correlation coefficient of $-0.37$ with a $p$-value of $0.05$. These values are also presented in Table \ref{tab:SR_res}.

\begin{figure}[h!]
    \centering
    \includegraphics[width = 1 \linewidth]{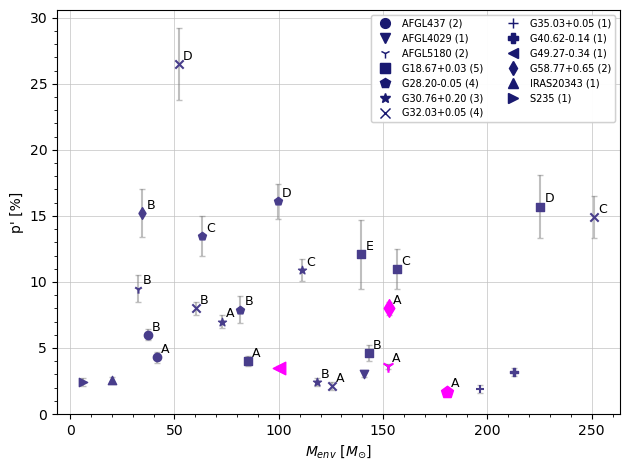}
    \caption{Polarization fraction on the SOMA scale against envelope mass from  Table \ref{tab:SED_params}. The four sources we have outflow direction for are plotted with larger, magenta symbols.}
    \label{fig:p_v_M}
\end{figure}

We also examine the relationship between the polarization fraction and the other protostellar properties from Table \ref{tab:SED_params}, i.e., envelope mass $M_{\rm env}$, mass surface density of the clump environment $\Sigma_{\rm cl}$, and mass of the star $m_\star$ in Figures~\ref{fig:p_v_M} and \ref{fig:p_v_sigma_mstar}. Neither show any significant correlation, with Spearman Rank statistics and $p$-values shown in Table \ref{tab:SR_res}.

\begin{table}[h!]
    \centering
    \caption{Spearman Rank statistics and corresponding $p$-value for the correlation between debiased polarization fraction and protostellar properties. }
    \begin{tabular}{lcc}
    \hline\hline
         Property & SR-stat. & $p$-value\\
         \hline
         $L_{\rm bol}$ & $-0.199$ & $0.309$ \\
         $M_{\rm env}$ & $-0.081$ & $0.683$ \\
         $\Sigma_{\rm cl}$ & $0.250$ & $0.200$ \\
         $m_\star$ & $-0.237$ & $0.225$\\
         $L_{\rm bol}/M_{\rm env}$ & $-0.371$ & $0.052$ \\
         \hline
    \end{tabular}
    \label{tab:SR_res}
\end{table}

\subsection{Relative orientation between magnetic field and outflow orientation}

\begin{figure*}
    \centering
    \includegraphics[width=0.9\linewidth]{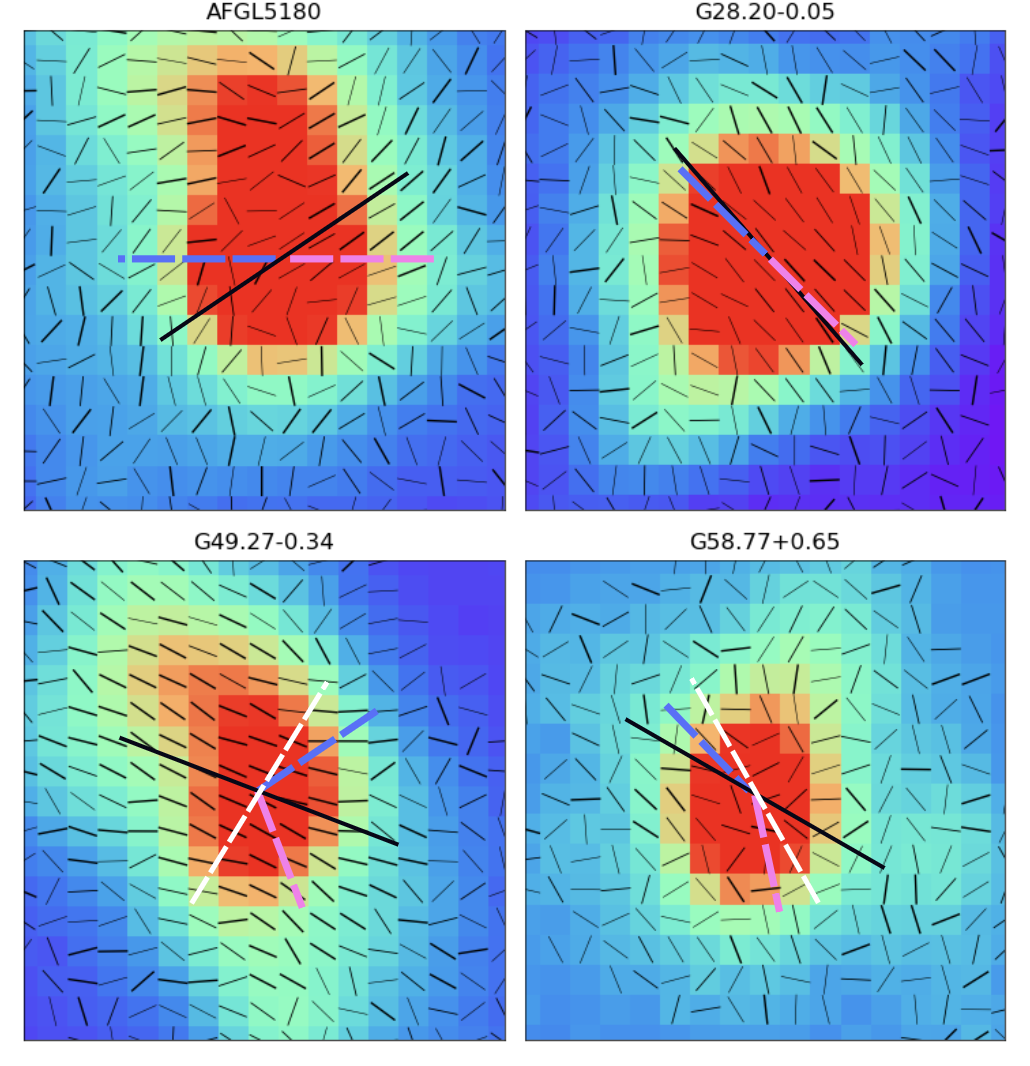}
    \caption{The figure illustrates the $\theta_{\rm Outflow}$ for the four sources. The blue dashed line shows the blue-shifted outflow orientation. The pink dashed line is the red-shifted outflow, and the white line is the average outflow orientation. The black line is the mean magnetic field orientation within the SOMA aperture for each source. The smaller black lines shows the magnetic field, with thicker lines having greater signal to noise ratio.  }
    \label{fig:outflow}
\end{figure*}

The averaged outflow orientation ($\theta_{\rm Outflow}$) is known for four of the sources studied in this work, that is for AFGL 5180 A \citep{crowe2024near}, G28.20 A \citep{law2022isolated}, and G49.27, G58.77 A (Rahman et al. in prep.). The red- and blue-shifted outflow directions for these sources, as well as the magnetic field orientations at the SOMA-scale are displayed in Table \ref{tab:outflow} and visualized in Figure \ref{fig:outflow}. 

\setlength{\tabcolsep}{1pt}
\begin{table}[h!]
    \centering
    \caption{Red- and blue shifted outflow in degrees for four sources. The magnetic field orientation from the sources, taken from table \ref{tab:basicprops}, and rounded to two significant figures are also included, as well as the relative orientation between magnetic field and outflow. All values are in degrees, normalized to be in the interval $[-90^\circ, 90^\circ]$, the relative orientation is normalized to the interval $[0, 90^\circ]$.}
    \begin{tabular}{lcccccc}
    \hline\hline
        \textbf{Source} & B-field or.  [$^\circ$] & \multicolumn{2}{c}{Outflow or. [$^\circ$]} & &\multicolumn{2}{c}{Relative or. [$^\circ$]}\\
        \cline{3-4} \cline{6-7} & &  Blue & Red & &  Blue & Red \\
    \hline
    AFGL 5180 A & $-56$ & $90$ & $-90$ & &$34$ & $34$\\
    \hline
    G28.8 A & $41$ & $45$ & $45$ & & $4$ & $4$\\
    \hline
    G49.27 & $69$ & $-85$ & $22$ & & $26$ & $47$\\
    \hline 
    G58.77 A & $61$ & $44$ & $15$ & & $17$ & $46$\\
    \hline
    \end{tabular}
    \label{tab:outflow}
\end{table}

As the $\theta_{\rm Outflow}$ is known for only four of the studied sources, the sample is too small to draw any conclusions on the link between outflow orientation and the mean magnetic field orientation. G28.2 A has the best alignment between the averaged magnetic field orientation and the outflow axis. We also know that this source is relatively isolated \citep{law2022isolated}. The rest of these sources do not align with the outflow in the same way, which may be related to the fact that they are in more clustered regions \citep{telkamp2025sofia}. Larger samples of sources with well measured outflow axes are required for more quantitative analysis.


\section{Conclusions} 
\label{sec:conclusion}

We have conducted a statistical study of the dust polarization and magnetic field properties of 29 SOMA sources and their host filaments. The main findings are summarized below. 

\begin{itemize}
    \item For the high intensity regime, we obtained $p'$ versus $I$ indices of $\alpha \sim0.6-0.8$, which suggests that dust grain alignment persists in these environments. 
    \item We have found a statistically significant bimodal distribution ($p_{\rm Kuiper}\lessapprox 0.05$) between the SOMA source orientations and the local magnetic field orientations. This suggests that magnetic fields play a dynamically important role in high-mass star-forming regions, similar to results reported for low-mass regions. 
    \item The bimodal distribution for the filament case is less clear, likely due to the inclusion of noisier data toward the outer part of the filament, making the filament orientation less well defined.
    \item We have not found correlations in $B-$field orientation between pairs of neighboring sources, which differs from some previous studies of star-forming regions \citep{li2014link}. This may indicate a scale at which gravitationally induced motions controlling orbits in the protocluster potential begin to dominate over large scale magnetic field dynamics.
\end{itemize}


\begin{acknowledgements}
  T.K. acknowledges support from a Chalmers Astrophysics and Space Sciences Summer (CASSUM) research fellowship. C.Y.L acknowledges financial support through the INAF Large Grant The role of MAGnetic fields in MAssive star formation (MAGMA). J.C.T. acknowledges ERC Advanced Grant MSTAR 788829 and NSF AST grants 2009674 and 2206450. K.P. is a Royal Society University Research Fellow, supported by grant number URF\textbackslash R1\textbackslash 211322. These observations were obtained by the James Clerk Maxwell Telescope, operated by the East Asian Observatory on behalf of The National Astronomical Observatory of Japan; Academia Sinica Institute of Astronomy and Astrophysics; the Korea Astronomy and Space Science Institute; the National Astronomical Research Institute of Thailand; the Center for Astronomical Mega-Science (as well as the National Key R\&D Program of China with No. 2017YFA0402700). Additional funding support is provided by the Science and Technology Facilities Council of the United Kingdom and participating universities and organizations in the United Kingdom and Canada.  Additional funds for the construction of SCUBA-2 were provided by the Canada Foundation for Innovation. The authors wish to recognize and acknowledge the very significant cultural role and reverence that the summit of Maunakea has always had within the indigenous Hawaiian community.  We are most fortunate to have the opportunity to conduct observations from this mountain. 
\end{acknowledgements}

\bibliographystyle{aa} 
\bibliography{references}

\begin{appendix} 
\onecolumn
\section{Grid Size of Stokes Parameter Maps}
In this work we have selected to carry out the analysis at a pixel scale of $4^{\prime\prime}$. To ensure that the grid size of the Stokes parameters maps does not affect the results, the basic properties in Table \ref{tab:basicprops} was re-calculated at a $12"$ pixel scale toward $G28.20 A$ as an example. We show that the result does not change significantly when using different gird sizes. The values for magnetic field orientation (weighted), dispersion and polarization fraction (weighted, debiased) are provided for both grid sizes on both beam and SOMA scale in Table \ref{tab:grids}. 

\setlength{\tabcolsep}{5pt}
\begin{table}[h!]
    \centering
    \caption{Basic properties, including magnetic field orientation, dispersion and debiased polarization fraction on beam, and SOMA scale derived from both the $4"$, and $12"$ grids for G28.2 $A$. The magnetic field orientation and debiased polarization fraction are both the intensity weighted averages inside the apertures.}
    \begin{tabular}{ccccccc}
    \hline
        \textbf{Grid size} & $\mathbf{\Bar{\theta}_{Beam}}$ & $\mathbf{\Bar{\theta}_{SOMA}}$ & $\mathbf{D_{Beam}}$ & $\mathbf{D_{SOMA}}$ & $\mathbf{\Bar{p'}_{Beam}}$ & $\mathbf{\Bar{p'}_{SOMA}}$\\
        (arcsec) & (deg) & (deg) & (deg) & (deg) & (\%) & (\%)\\
    \hline
        4 & 41.20 $\pm$ 2.10 & 40.9 $\pm$ 1.90 & 24.99 $\pm$ 2.56 & 24.03 $\pm$ 2.15 & 1.00 $\pm$ 0.30 & 1.70 $\pm$ 0.20\\
        12 & 48.90 $\pm$ 2.50 & 47.00 $\pm$ 1.20 & 33.16 $\pm$ 20.11 & 23.82 $\pm$ 18.76 & 0.70 $\pm$ 0.10 & 1.00 $\pm$ 0.00\\ 
    \hline
    \end{tabular}
    \label{tab:grids}
\end{table}

\begin{figure}[h!]
    \centering
    \includegraphics[width=0.8\linewidth]{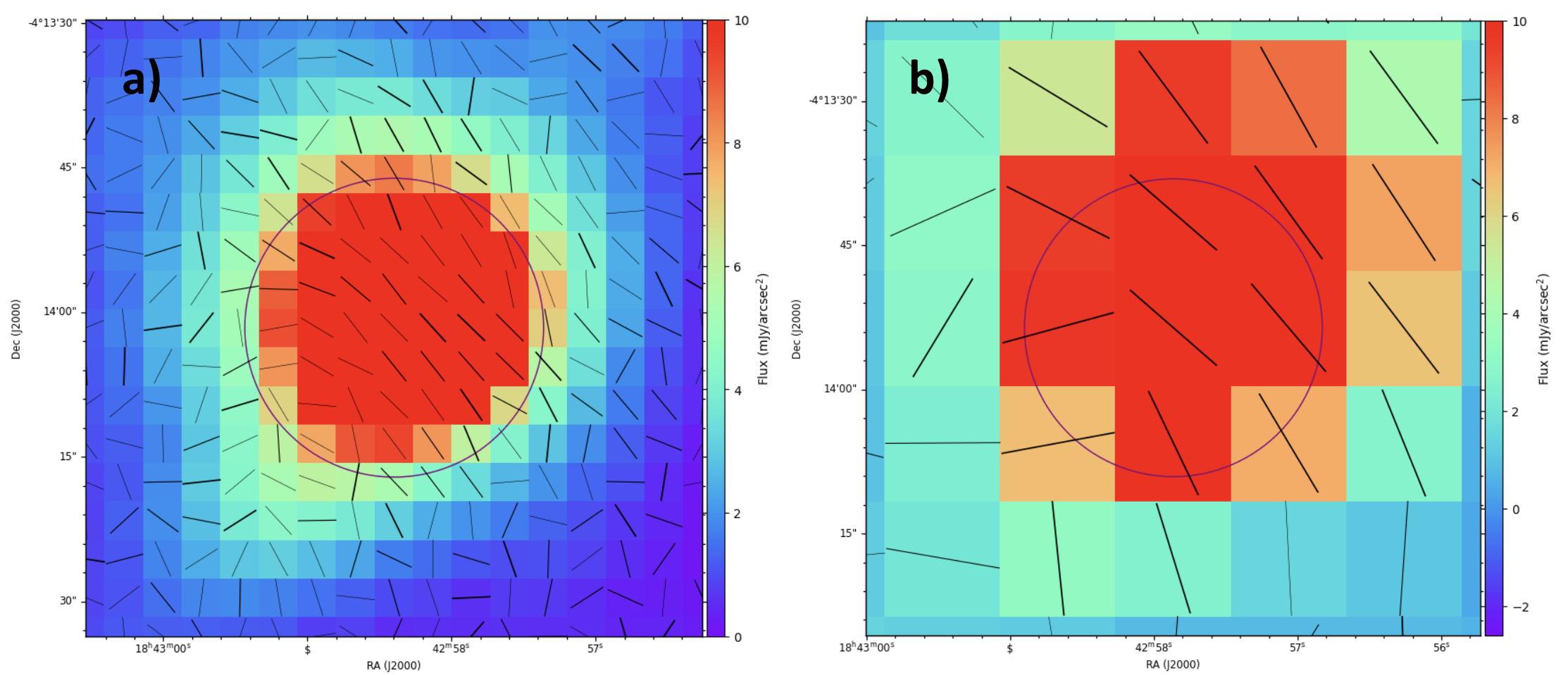}
    \caption{Source \textit{G28.20 A} including magnetic field vectors and SOMA aperture on both the $4"$ grid (a) and the $12"$ grid (b). The thickness of the vectors has equivalent meaning as in Figure 3. }
    \label{fig:G28.2_grirds}
\end{figure}

Figure \ref{fig:G28.2_grirds} shows $G28.20 A$ on both the $4"$ and $12"$ grid. 

\begin{figure}[h!]
    \centering
    \includegraphics[width=0.8\linewidth]{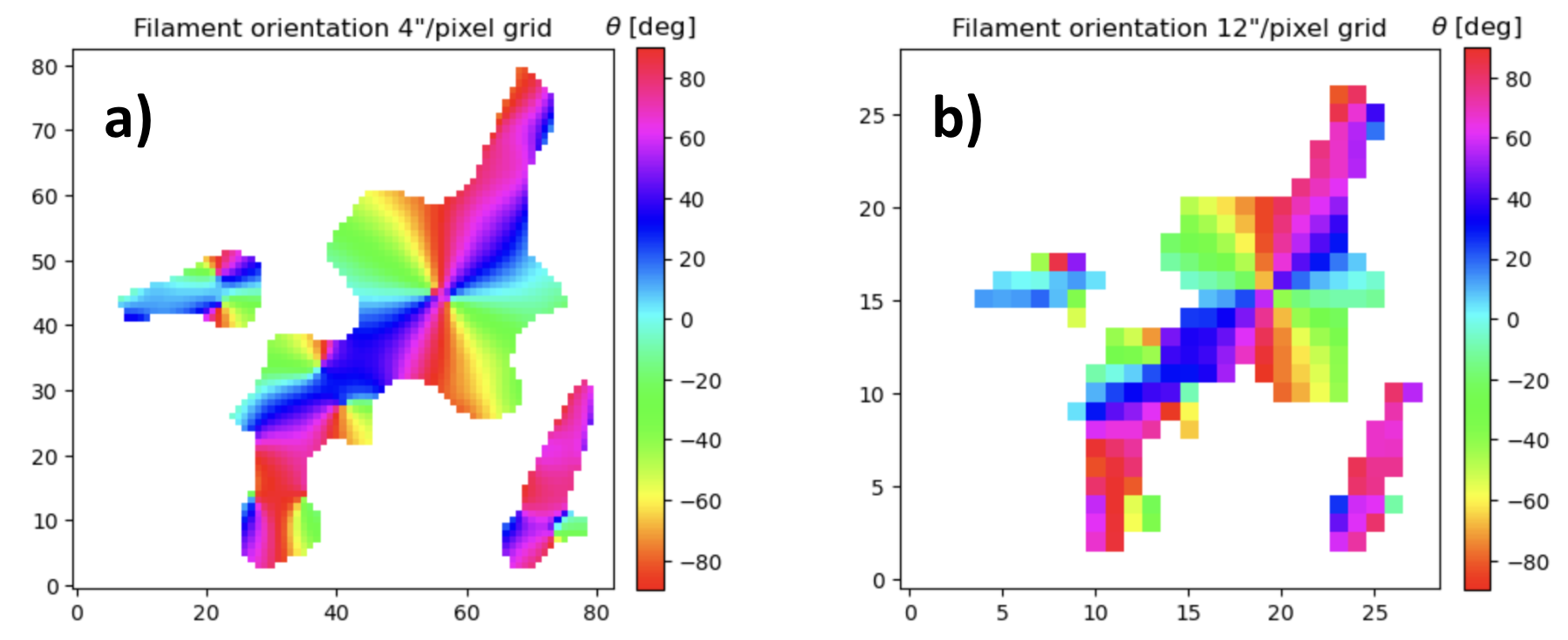}
    \caption{The filament of \textit{G28.20 A} identified from the Hessian Matrix analysis, on the $4"$ grid (a) and the $12"$ grid (b). The orientation of the filament is also colorcoded into the figure. }
    \label{fig:hessian_grids}
\end{figure}

\begin{figure}[h!]
    \centering
    \includegraphics[width=0.8\linewidth]{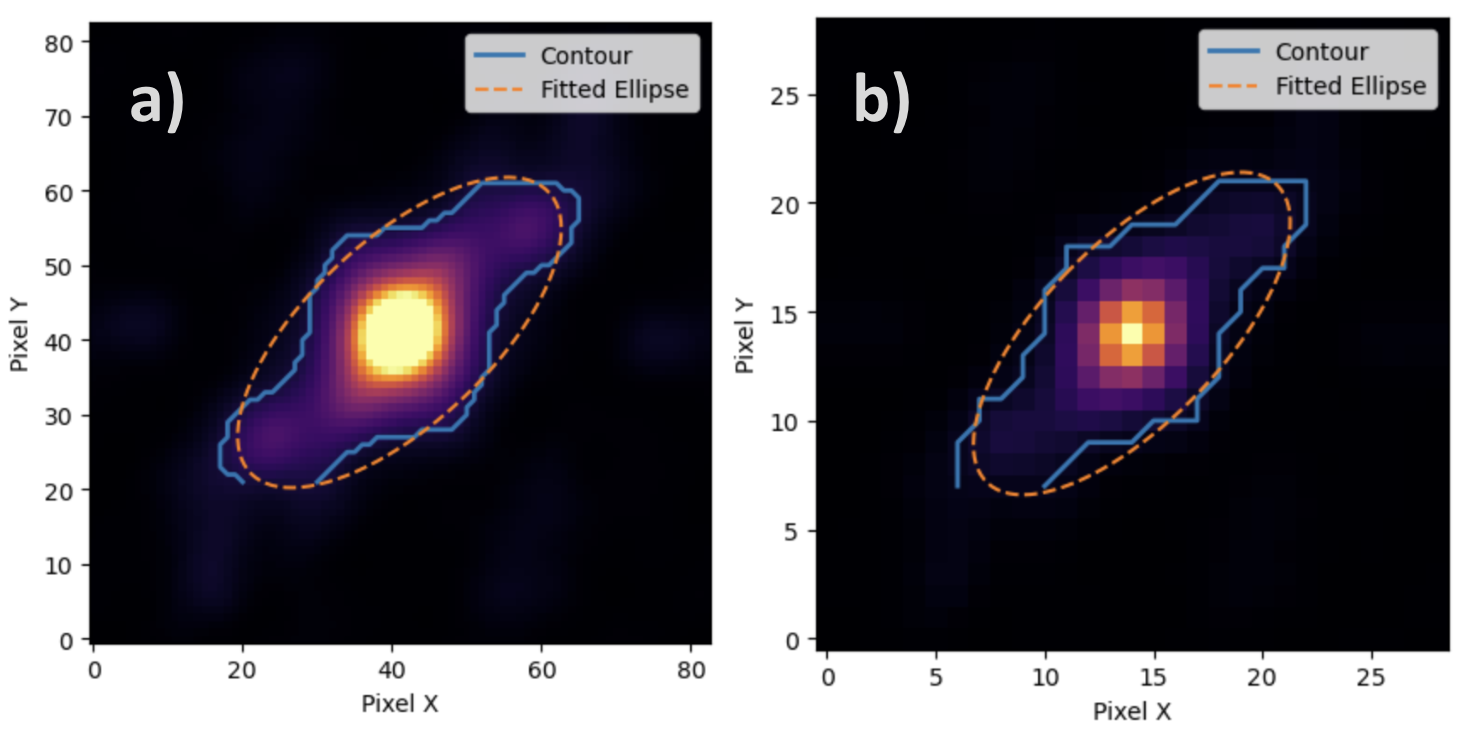}
    \caption{The autocorrelation function of Stokes I map of the filament of \textit{G28.20 A} identified from the Hessian Matrix analysis (see Figure \ref{fig:hessian_grids}) on the $4"$ grid (a) and the $12"$ grid (b). The orange dashed line shows the ellipse fitted to the blue contour, from which the orientation of the filament was obtained.  }
    \label{fig:autocorr_grids}
\end{figure}

We also see from Figures \ref{fig:hessian_grids} -- \ref{fig:autocorr_grids} that the estimated filament orientation is likely not affected by the used grid size, for either of the methods. Figure \ref{fig:hessian_grids} shows the filament of $G28.20 A$ identified using Hessian matrix analysis for the two grid sizes, and Figure \ref{fig:autocorr_grids} shows the autocorrelation function and fitted ellipse used to derive the orientation. 

\end{appendix}

\end{document}